\documentclass[
 preprint,
 superscriptaddress,
 amsmath,amssymb,
 aip,
]{revtex4-1}

\usepackage{graphicx}
\usepackage{dcolumn}
\usepackage{bm,bbm}
\usepackage{xcolor}
\usepackage{comment}

\newcommand{\Wt}{\tilde{W}}
\newcommand{\eps}{\varepsilon}
\newcommand{\ii}{\mathrm{i}}

\begin{document}
\title{Real-Time Approach to the Dynamical Bethe-Salpeter Equation for Finite Systems}
\author{Tucker Allen}
\email{tuckerallen27@ucla.edu}
\affiliation{Department of Chemistry and Biochemistry,
University of California, Los Angeles, California 90095, USA}

\author{Barry Y. Li}
\affiliation{Department of Chemical and Biological Physics,
Weizmann Institute of Science, Rehovot 7610001, Israel}

\author{Daniel Neuhauser}
\affiliation{Department of Chemistry and Biochemistry and California Nanoscience Institute,
University of California, Los Angeles, California 90095, USA}

\date{\today}

\begin{abstract}
 
We present a real-time linear-response approach to solving the dynamical Bethe-Salpeter equation (BSE). The polarization part of the screened interaction is obtained from time-dependent Hartree propagation of the one-particle density matrix within an orbital basis-set representation. The frequency convolution defining the dynamical kernel is evaluated as a product in time, avoiding numerical integration and storage of the full screened Coulomb operator. The resulting nonlinear eigenvalue problem is then solved directly, going beyond the static screening approximation with full-frequency dependence in the screened interaction. We validate the method by calculating the low-lying singlet and triplet excitation energies of water, formaldehyde, and acetylene. While the deterministic scaling remains $\mathcal{O}(N^6)$, the time propagation formulation is readily compatible with grid-based stochastic sampling methods, which will open the possibility for dynamical BSE calculations of very large systems.
\end{abstract}

\maketitle

\section{Introduction}

Developing computationally efficient methods that accurately describe molecular excited states remains an active area of research. Coupled-cluster and multiconfigurational wave-function methods can provide very accurate results for small molecules, but their computational scaling limits their applicability to large systems. Time-dependent density-functional theory (TDDFT) is considerably less expensive and is widely used for first-principles excited-state calculations. However, the standard adiabatic approximation neglects memory effects by assuming that the exchange-correlation kernel is local in time.\cite{ullrich_snapshot_2025} Consequently, standard adiabatic exchange-correlation kernels can struggle to describe excitation processes that depend explicitly on frequency, including states with substantial multiple-excitation character.\cite{Maitra2004,levine_conical_2006,Romaniello2009,Sangalli2011,Zhang2013,Rebolini2016,Lacombe2023}

Green's-function-based many-body perturbation theory provides an alternative route to molecular excited states.\cite{Hedin_1965,Louie1985,Strinati,Blase2020} Within this framework, one begins from a mean-field ground-state description often obtained with density-functional or Hartree-Fock theory. Quasiparticle spectra, ionization potentials, and electron affinities can be calculated within the $GW$ approximation, while neutral optical excitations are obtained by solving the two-particle Bethe-Salpeter equation (BSE). The BSE includes excitonic effects through an explicit electron-hole interaction kernel. The combined $GW$-BSE approach has become increasingly popular for studying molecular systems with its ability to describe valence, charge-transfer, and Rydberg excitations more consistently than TDDFT, even rivaling CCSD and CASPT2 methods.\cite{Hedin_1965,Strinati,Louie_BSE_2000,Onida_RMP_2002,Blase2020,jacquemin_is_2017,li_bethe-salpeter_2022}

Most implementations of the BSE employ the static-screening approximation for the screened Coulomb interaction entering the electron-hole kernel. In this approximation, the frequency-dependent dynamical interaction $\tilde{W}(\Omega)$ is replaced by the static interaction
$W(\omega=0)$. This assumes that the electron-hole attraction is instantaneous. We emphasize that this approximation applies specifically to the screened interaction entering the BSE kernel; the frequency dependence of $W$ is generally retained in the preceding $GW$ calculation of the quasiparticle self-energy. This approximation has enabled a large body of successful calculations for a wide variety of systems.\cite{Louie1985,Louie_BSE_2000,Rocca2012-WEST,Jacquemin2015,Gui2018,forster_quasiparticle_2022,Blase2020} The static approximation is expected to be accurate when the screening response is much faster than the electron-hole dynamics. This translates to excitation energies that are well separated from the characteristic screening modes, i.e., plasmons, of the system. Dynamical effects can become important when this separation of energy scales breaks down, as demonstrated for copper and silver metals in Ref. \cite{marini_dynamical_2003}.

The static approximation nevertheless represents a physical restriction because it removes the delay, or memory, in the polarization response of the electronic system. This restriction can be especially problematic in finite systems.\cite{Louie_BSE_2000,Bechstedt2015} Molecules generally exhibit weak screening and can have large exciton binding energies, such that the poles of $W(\omega)$ correspond to molecular polarization modes that need not be energetically well separated from the optical excitation energies. In this regime, dynamical screening may not be adequately represented as a small correction to a static interaction. The static approximation can result in errors in energies on the order of $0.1-0.3$ eV.\cite{Ma2009} In general, the static $GW$-BSE performs best for spin-conserved transitions, with a poor description of triplet states.\cite{Jacquemin2015,jacquemin_is_2017}

Several approaches have been developed to incorporate frequency dependence into the BSE. Perturbative dynamical corrections evaluate the frequency-dependent interaction using eigenstates obtained from a static BSE calculation.\cite{Louie1985,Loos2020dyn} Various plasmon-pole models have been developed for both solids and molecules.\cite{Louie_BSE_2000,Ma2009,Zhang2023,Wen2026,Zhou2026} An alternative formulation is to convert the frequency dependence into a larger static eigenvalue problem.\cite{Bintrim2022} These studies have demonstrated that dynamical screening tends to redshift singlet and triplet excitation energies and generally improves numerical agreement with high-level quantum chemical benchmarks.\cite{Loos2020dyn,Bintrim2022,Wen2026}

In principle, a frequency-dependent kernel can also introduce additional poles in the response function. This property provides a possible route to states with double- or higher-excitation character that are absent from the single-particle transition space of a static BSE or adiabatic TDDFT calculation.\cite{Maitra2004,Kurzweil2008,Lacombe2023,Romaniello2009,Sangalli2011,daas_prl_2026} Such effects are relevant in systems with open-shell ground states as well as excited states with double-excitation character.\cite{Romaniello2009,casanova_spin-flip_2020,Park2021,Dar2025}

Practical solutions of the nonlinear dynamical BSE remain computationally demanding because the screened interaction must be constructed and the BSE matrix diagonalized over many sampled frequencies.\cite{Romaniello2009,Sangalli2011,Zhang2013,Rebolini2016,Authier2020,Loos2022,Bintrim2022} In a conventional grid-based frequency-domain implementation, the independent-particle irreducible polarizability $\chi_0(\omega)$ is evaluated through a sum-over-states expression and the random phase approximation (RPA) dielectric matrix is inverted to obtain $W(\omega)$. In general, the explicit evaluation and storage of the screened Coulomb operator over a dense frequency grid can become a major CPU and memory bottleneck. We do note that various efficient methods that avoid dielectric matrix inversion for $GW$ \cite{PhysRevB.78.113303,ren_resolution--identity_2012,neuhauser2014breaking,duchemin_cubic-scaling_2021,yeh_fully_2022} and static $GW$-BSE \cite{Rocca2012-WEST,Rabani2015,bradbury_optimized_2023,hillenbrand_energy-specific_2025} have been developed. 

In this work, we present a real-time linear-response method for building the frequency-dependent interaction kernel entering the dynamical BSE for finite systems. The retarded reducible RPA response is obtained by propagating the time-dependent Hartree (TDH) equations following impulsive perturbations generated from orbital pair densities. The resulting induced density is contracted with the Coulomb tensor to obtain the required screened-interaction matrix elements. After constructing the corresponding time-ordered interaction, the convolution defining $\tilde{W}(\Omega)$ is evaluated as a product in the time domain followed by a Fourier transform. This avoids explicit numerical integration over frequency and does not require storage of the full real-space screened Coulomb operator. 

We validate the real-time formalism by calculating the low-lying singlet and triplet excitation energies of $\rm{H_2O}$, $\rm{CH_2O}$, and $\rm{C_2H_2}$. These molecules provide representative Rydberg, $n\to\pi^*$, and $\pi\to\pi^*$ excitations, respectively. The resulting static excitation energies and dynamical corrections are compared with those reported by Loos and Blase using a renormalized first-order perturbative treatment of dynamical screening.\cite{Loos2020dyn}

The present formulation is the dynamical extension of our earlier stochastic real-time approaches for representing the screened Coulomb matrix in large molecular systems.\cite{neuhauser2014breaking,Rabani2015,vlcek_swift_2018,bradbury_bethesalpeter_2022,bradbury_optimized_2023,no_more_gap,Allen2026} Although the deterministic orbital-basis implementation employed here retains the formal $\mathcal{O}(N^6)$ scaling, the real-time construction is directly compatible with stochastic sampling and grid-based propagation. 

The paper is organized as follows. We first present the theoretical formulation and real-time implementation. We then demonstrate the method for low-lying singlet and triplet excitations in small molecules and compare the direct nonlinear solutions against previously reported first-order dynamical corrections.\cite{Loos2020dyn} We conclude with a discussion of future directions and the extension of the real-time approach to much larger systems.

\section{Theory}
In this section, we develop the real-time method for evaluating the dynamically screened Coulomb interaction, $\tilde{W}(\Omega)$. Both the software implementation and results presented here employ the Tamm-Dancoff Approximation (TDA) to the excitonic BSE Hamiltonian. The resonant-antiresonant coupling elements of $\tilde{W}$ can produce numerical divergences \cite{Loos2020dyn} and will be explored in future work.

\subsection{Dynamical BSE}
The BSE is cast as a nonlinear eigenvalue problem in the space of $N_vN_c$ valence-to-conduction transitions: 
\begin{equation}
    \tilde{A}_{ia,jb}(\Omega)=(\eps_a^{\rm{GW}}-\eps_i^{\rm{GW}})\delta_{ij}\delta_{ab}+\kappa(ia|jb)-\tilde{W}_{ab,ij}(\Omega),
    \label{eq:tda_dbse_matrix}
\end{equation}
where $\varepsilon_i^{\rm{GW}}$ are occupied (valence) and $\varepsilon_a^{{\rm{GW}}}$ unoccupied (conduction) {$GW$} energies. Here, we apply one-shot diagonal $G_0W_0$ energy corrections on top of eigenvalues of a restricted Hartree-Fock calculation. The second term is the repulsive exciton exchange interaction responsible for singlet-triplet splitting, with $\kappa=0,2$ for triplet and singlet excitations, respectively. Two-electron Coulomb integrals $(pq|rs)$ are expressed in $(rr|r'r')$ notation: 
\begin{equation}
    (pq|rs)=\int dr'dr\, \phi_p(r)\phi_q(r)|r-r'|^{-1}\phi_r(r')\phi_s(r'),
\end{equation}
where the Hartree-Fock orbitals are taken to be real.
Throughout the manuscript, occupied molecular orbitals (MOs) are labeled $i,j,k,l$, virtual MOs by $a,b,c,d$, and general MOs by $p,q,r,s$. The direct electron-hole interaction has matrix elements: 
\begin{equation}
\begin{split}
\tilde{W}_{ab,ij}(\Omega)=\frac{\ii}{2\pi}\int  d\omega\, W_{ab,ij}(\omega) \times
\Bigg[&
\frac{1}{\Omega-(\eps_b^{\rm{GW}}-\eps_i^{\rm{GW}}+\omega)+\ii\eta} \\
&+\frac{1}{\Omega-(\eps_a^{\rm{GW}}-\eps_j^{\rm{GW}}-\omega)+\ii\eta}
\Bigg],
\end{split}
\label{eq:wtilde_frequency}
\end{equation}
where $ W_{ab,ij}(\omega)= \int dr'dr\,\phi_a(r)\phi_b(r)W(r,r',\omega)\phi_i(r')\phi_j(r')$, and $\eta$ is the numerical broadening. Eq. (\ref{eq:wtilde_frequency}) is a Shindo-like approximation, where the frequency dependence of the BSE is factorized separately into the screened interaction and non-interacting electron-hole propagators.\cite{Strinati,Romaniello2009,Bechstedt2015,Zhang2023} A derivation of Eq. (\ref{eq:tda_dbse_matrix}) is provided in Appendix \ref{app:shindo_dbse_derivation}. While calculating Eq.  (\ref{eq:tda_dbse_matrix}) is relatively straightforward, it is numerically very expensive to build the real-space $W(r,r',\omega)$ operator over many frequencies and then perform the $\mathcal{O}(N_\Omega N_\omega)$ numerical integration over $\mathcal{O}(N_v^2 N_c^2)$ electron-hole matrix elements in Eq. (\ref{eq:wtilde_frequency}). We note that an alternative to constructing $W(\omega)$ on a frequency grid is to obtain its spectral representation by diagonalizing the direct RPA matrix, although this diagonalization scales as $\mathcal{O}(N^6)$.\cite{Loos2020dyn,sun_recent_2020}

Solution of the nonlinear eigenvalue problem:
\begin{equation}
    \tilde{A}(\Omega)X_n(\Omega)=\lambda_n(\Omega)X_n(\Omega),
    \label{eq:nonlinear_root}
\end{equation}
yields physical excitation energies that satisfy $\mathrm{Re}\ \lambda_n(\Omega)=\Omega$. We obtain the roots of Eq. (\ref{eq:nonlinear_root}) by a linear interpolation method, which we discuss later. The standard static BSE is obtained by replacing $\tilde{W}_{ab,ij}(\Omega)\approx W_{ab,ij}(\omega=0)$, reducing the nonlinear eigenproblem to a linear one. In this work, we present results for both the static and dynamical BSE. 

\subsection{Real-Time Based Approach to $\tilde{W}_{ab,ij}(\Omega)$}

We outline how the electron-hole kernel is constructed through time-propagation. Rather than explicitly performing the frequency convolution in Eq. (\ref{eq:wtilde_frequency}), we calculate the polarization part of the screened Coulomb interaction through real-time linear response theory and evaluate the frequency convolution as a product in the time domain.

It is useful to first separate the time-ordered screened Coulomb interaction into its instantaneous and polarization parts,
\begin{equation}
W_{ab,ij}(t)=(ab|ij)\delta(t)+W^{\rm{pol}}_{ab,ij}(t).
\label{eq:w_decomposition}
\end{equation}
The real-time propagation is used to obtain the reducible RPA polarization function \cite{neuhauser2014breaking,cunningham_many-body_2025}, from which the polarization part of the screened interaction is constructed, while the instantaneous bare Coulomb term is added separately.

The real-time approach begins with a finite orbital basis. We expand the time-dependent Hartree-Fock orbitals as
\begin{equation}
\psi_p(r,t)=\sum_s C_{sp}(t)\phi_s(r),
\label{eq:ct_expansion}
\end{equation}
where the coefficient matrix includes both occupied and unoccupied orbitals. For simplicity, we consider a single perturbing occupied-orbital pair $kl$, with the TDA BSE containing a total of $N_v(N_v+1)/2$ unique pairs. The corresponding perturbing potential produced by $kl$ is represented in the orbital basis by
\begin{equation}
D_{pq}^{kl}=(pq|kl),
\label{eq:source_matrix}
\end{equation}
and an impulsive perturbation of strength $\lambda$ is applied:
\begin{equation}
C^{kl}(t=0^+)=e^{-\ii\lambda D^{kl}}C(t=0),
\label{eq:kick_exact}
\end{equation}
where $\lambda$ is taken to be $0.0001$ $E_h^{-1}$. Here, we diagonalize each $D^{kl}$; however, one could also use a normalized first-order Taylor expansion, i.e., $C^{kl}(t=0^+)\approx \mathcal{N}(1-\ii \lambda D^{kl})C(t=0)$, where $\mathcal{N}$ is a normalization factor with a correction quadratic in $\lambda$.

The perturbed coefficient matrix is propagated according to
\begin{equation}
\ii\frac{\partial C^{kl}(t)}{\partial t}
=
H^{kl}(t)C^{kl}(t),
\label{eq:tdh_equation}
\end{equation}
where the density matrix is
\begin{equation}
P_{pq}^{kl}(t)
=
\sum_m C_{pm}^{kl}(t)f_m C_{qm}^{kl,*}(t),
\label{eq:density_matrix}
\end{equation}
where $f_m=1$ for occupied orbitals and $f_m=0$ for unoccupied orbitals. To obtain the screened Coulomb interaction at the RPA level, we use the time-dependent Hartree Hamiltonian,
\begin{equation}
H_{pq}^{kl}(t)
= \eps_p\delta_{pq}+2\sum_{rs}(pq|rs)\left[
P_{rs}^{kl}(t)-P_{rs}(t=0)\right],
\label{eq:tdh_hamiltonian}
\end{equation}
where $\{\varepsilon_p\}$ are Hartree-Fock eigenvalues and the factor of two accounts for spin. The induced density matrix associated with the perturbing source $kl$ is
\begin{equation}
\delta P_{pq}^{kl}(t)
= \frac{1}{\lambda}[P_{pq}^{kl}(t)-P_{pq}(t=0)].
\label{eq:delta_p_source}
\end{equation}
The resulting retarded polarization contribution to the screened interaction is
\begin{equation}
W_{pq}^{{\rm{pol}},kl,R}(t)
=
2\sum_{rs}(pq|rs)\delta P_{rs}^{kl}(t).
\label{eq:wret_from_density}
\end{equation}
The causal frequency-domain response is obtained by applying a Gaussian damping and Fourier transforming the positive-time response,
\begin{equation}
W_{pq}^{{\rm{pol}},kl,R}(\omega)
=
\int_0^\infty d t\,
e^{-\gamma^2t^2/2}
e^{\ii\omega t}
W_{pq}^{{\rm{pol}},kl,R}(t).
\label{eq:wret_fft}
\end{equation}
For real-valued perturbations, the time-ordered and retarded interactions are related in frequency space by \cite{fetter_quantum_2012}
\begin{equation}
W_{pq}^{{\rm pol},kl}(\omega)
=
\begin{cases}
W_{pq}^{{\rm pol},kl,R}(\omega), 
& \omega \geq 0, \\[2mm]
\left[W_{pq}^{{\rm pol},kl,R}(\omega)\right]^*, 
& \omega < 0.
\end{cases}
\label{eq:retarded_to_timeordered}
\end{equation}
The time-ordered interaction on the full time axis is then obtained from
\begin{equation}
W_{pq}^{{\rm{pol}},kl}(t)
=
\frac{1}{2\pi}
\int_{-\infty}^{\infty}d\omega\,
e^{-\ii\omega t}
W_{pq}^{{\rm{pol}},kl}(\omega).
\label{eq:w_time_ordered_inverse}
\end{equation}
This causal-to-time-ordered transformation follows the real-time prescription used previously in stochastic $GW$ calculations.\cite{neuhauser2014breaking,vlcek_swift_2018} 

For a TDA BSE matrix element $\tilde{W}_{ab,ij}(\Omega)$, the occupied-occupied perturbing source is $kl=ij$ and the response is measured in the unoccupied-unoccupied sector $pq=ab$. The polarization contribution to the dynamical interaction can then be evaluated directly in the time domain as 
\begin{equation}
\begin{split}
\tilde{W}^{\rm{pol}}_{ab,ij}(\Omega)
=&
\int_0^\infty d t\,
W_{ab}^{{\rm{pol}},ij}(t)
e^{-\ii(\eps_b^{\rm{GW}}-\eps_i^{\rm{GW}})t}
e^{\ii\Omega t}
e^{-\eta|t|}
\\
&+
\int_{-\infty}^{0}d t\,
W_{ab}^{{\rm{pol}},ij}(t)
e^{\ii(\eps_a^{\rm{GW}}-\eps_j^{\rm{GW}})t}
e^{-\ii\Omega t}
e^{-\eta|t|}.
\end{split}
\label{eq:wtilde_time}
\end{equation}
Eq. (\ref{eq:wtilde_time}) is the central real-time expression used in our implementation. The orbital-energy phase factors originate from the energy denominators in Eq. (\ref{eq:wtilde_frequency}). The frequency convolution over $\omega$ is therefore replaced by multiplication in the time domain followed by Fourier transformation.

The full dynamically screened interaction entering the BSE is obtained by adding the instantaneous bare Coulomb contribution,
\begin{equation}
\tilde{W}_{ab,ij}(\Omega)
=
(ab|ij)+\tilde{W}^{\rm{pol}}_{ab,ij}(\Omega).
\label{eq:wtilde_full}
\end{equation}

The same real-time propagation also gives the zero-frequency screened interaction used for the static BSE,
\begin{equation}
W_{ab,ij}(\omega=0)
=
(ab|ij)
+
\int_0^\infty d t\,
e^{-\gamma^2t^2/2}
W_{ab}^{{\rm{pol}},ij,R}(t).
\label{eq:static_from_rt}
\end{equation}
The above procedure is repeated for each unique occupied-orbital pair. This task is trivially parallelizable over separate MPI processes, so time propagation for different perturbations can be performed simultaneously. \cite{neuhauser2014breaking,vlcek_swift_2018,bradbury_bethesalpeter_2022,bradbury_optimized_2023}

\section{Validation of the Method}

We validate the real-time formalism for a series of small molecules: $\rm{H_2O}$, $\rm{CH_2O}$, and $\rm{C_2H_2}$. This benchmark includes transitions of different character, allowing us to quantify the magnitude of dynamical corrections for representative Rydberg, $n\rightarrow\pi^{*}$, and $\pi\rightarrow\pi^{*}$ excitations. Molecular geometries are taken from Ref.~\cite{Wen2026}. For consistency with previous dynamical BSE molecular benchmarks,\cite{Loos2020dyn,Wen2026} we employ a restricted Hartree-Fock starting point and apply $G_0W_0$ corrections to every MO energy. All-electron aug-cc-pVTZ basis sets are used for the self-consistent-field and $G_0W_0$ calculations, which are performed with the \textsc{PySCF} software package.\cite{sun_recent_2020} The frequency dependence of the screened Coulomb interaction entering the $G_0W_0$ self-energy is treated exactly within the RPA by diagonalizing the direct RPA matrix.

\begin{table}[t]
\caption{Computational details for the molecular benchmarks using the aug-cc-pVTZ basis set. The Hartree-Fock and $G_0W_0$ gaps are obtained from all-electron calculations performed with \textsc{PySCF}.\cite{sun_recent_2020} $N_v$ and $N_c$ are the numbers of valence occupied and conduction unoccupied orbitals, respectively, included in the TDH propagation and BSE solver.}
\label{tab:computational_details}
\begin{ruledtabular}
\begin{tabular}{lcccc}
Molecule & HF gap (eV) & $G_0W_0$ gap (eV) & $N_v$ & $N_c$ \\
\hline
H$_2$O     & 14.69 & 13.57 & 4 & 87  \\
CH$_2$O    & 12.77 & 11.99 & 6 & 130 \\
C$_2$H$_2$ & 12.03 & 12.32 & 5 & 131 \\
\end{tabular}
\end{ruledtabular}
\end{table}

For the subsequent TDH propagation and BSE calculations, we employ the frozen-core approximation by excluding the $1s$ orbitals of the carbon and oxygen atoms. All remaining occupied orbitals and all virtual orbitals are retained, as in Table~\ref{tab:computational_details}. In comparison to all-electron calculations, the frozen-core approximation changes the low-lying excitation energies by less than $5$ meV.

We first examine numerical convergence with respect to the maximum positive propagation time, $T_{\max}$. A time step of $dt=0.05$ a.u. is used throughout. At each value of $T_{\max}$, the Gaussian damping is set to $\gamma=3/T_{\max}$. Because the time-ordered interaction is defined over $[-T_{\max},T_{\max}]$, the Fourier-grid spacing is $\Delta\Omega=\pi/T_{\max}$, and we set the exponential broadening to $\eta=\Delta\Omega/2$. For the reported excitation energies, we use a $T_{\max}\approx 2.5$ fs, corresponding to $\gamma\approx 0.8$ eV and $\eta\approx 0.4$ eV.

For each molecule and propagation time, we diagonalize the frequency-dependent TDA-BSE matrix on a coarse set of 10--20 trial frequencies $\Omega$. The low-lying eigenvalue branches vary smoothly with $\Omega$, allowing the dynamical roots to be determined by linear interpolation between neighboring trial frequencies.\cite{Zhang2023} The physical excitation energies correspond to the intersections $\mathrm{Re}\,\lambda_n(\Omega)=\Omega$, as defined in Eq.~(\ref{eq:nonlinear_root}). Thus, only a small number of frequency-dependent BSE diagonalizations are required to determine the low-lying dynamical excitation energies.

\begin{figure*}[t]
\centering
\includegraphics[width=0.95\textwidth]{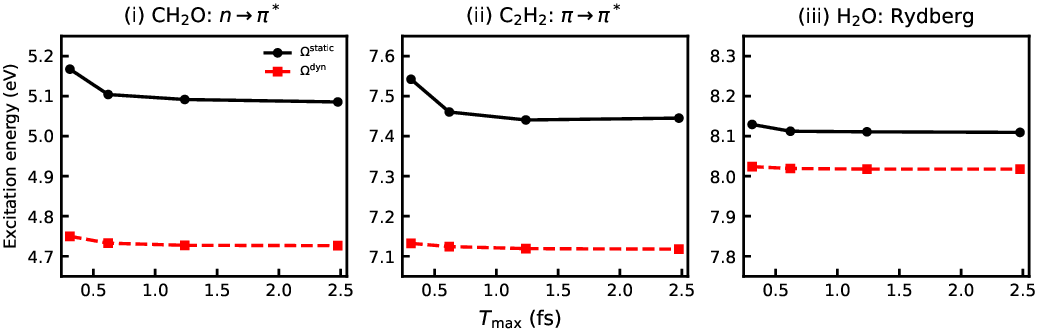}
\caption{Convergence of representative low-lying singlet excitation energies with maximum propagation time, $T_{\max}$: (i) the $^{1}A_2$ $n\rightarrow\pi^{*}$ excitation of CH$_2$O, (ii) the $^{1}\Sigma_u^{-}$ $\pi\rightarrow\pi^{*}$ excitation of C$_2$H$_2$, and (iii) the $^{1}B_1$ $n\rightarrow 3s$ Rydberg excitation of H$_2$O. Static and dynamical BSE excitation energies are shown as black circles and red squares, respectively.}
\label{fig:time_convergence}
\end{figure*}

Figure~\ref{fig:time_convergence} shows the convergence of various singlet excitations. Doubling the propagation time from $1.24$ to $2.48$ fs changes the dynamical excitation energies by at most $1.3$ meV, while the static excitation energies and dynamical shifts change by no more than $6$ meV. These results also indicate that the low-lying dynamical roots can be converged without requiring long-time propagation.

Tables~\ref{tab:h2o_excitations}--\ref{tab:c2h2_excitations} compare the static and dynamical excitation energies obtained in the present work with those reported by Loos and Blase.\cite{Loos2020dyn} The term symbols and excitation characters are assigned from the irreducible representations of the contributing electron-hole pairs and the dominant components of the static BSE eigenvectors. The present calculations employ the TDA for both the static and dynamical BSE, and the dynamical excitation energies are obtained by directly solving the nonlinear, frequency-dependent eigenvalue problem. In comparison, Ref.~\cite{Loos2020dyn} constructs the full RPA spectral representation of the screened Coulomb interaction and solves the full static BSE including resonant-antiresonant coupling. The dynamically screened Coulomb potential is then evaluated at the static excitation energy and incorporated through a renormalized first-order perturbative correction within the dynamical TDA. Because the static BSE treatments differ, the dynamical shift $\Delta=\Omega^{\mathrm{dyn}}-\Omega^{\mathrm{static}}$ provides the most direct comparison between the two treatments of frequency-dependent screening.

\begin{table*}[t]
\caption{Singlet and triplet BSE excitation energies of H$_2$O calculated with static ($\Omega^{\mathrm{static}}$) and dynamical ($\Omega^{\mathrm{dyn}}$) screening. All energies are in eV and $\Delta=\Omega^{\mathrm{dyn}}-\Omega^{\mathrm{static}}$. The reference values are from Ref.~\cite{Loos2020dyn}, where all-electron aug-cc-pVTZ dynamical BSE@$G_0W_0$@HF calculations are performed with dynamical screening included through a renormalized first-order perturbative correction.}
\label{tab:h2o_excitations}
\begin{ruledtabular}
\begin{tabular}{llcccccc}
& & \multicolumn{3}{c}{Present work} & \multicolumn{3}{c}{Ref.~\cite{Loos2020dyn}} \\
State
& Character
& $\Omega^{\mathrm{static}}$
& $\Omega^{\mathrm{dyn}}$
& $\Delta$
& $\Omega^{\mathrm{static}}$
& $\Omega^{\mathrm{dyn}}$
& $\Delta$ \\
\hline
$^{1}B_{1}$ & $n\rightarrow 3s$ & 8.11  & 8.02  & $-0.09$ & 8.09  & 8.00  & $-0.09$ \\
$^{1}A_{2}$ & $n\rightarrow 3p$ & 9.79  & 9.71  & $-0.08$ & 9.79  & 9.72  & $-0.07$ \\
$^{1}A_{1}$ & $n\rightarrow 3s$ & 10.44 & 10.36 & $-0.07$ & 10.42 & 10.35 & $-0.07$ \\
$^{3}B_{1}$ & $n\rightarrow 3s$ & 7.64  & 7.49  & $-0.15$ & 7.62  & 7.48  & $-0.14$ \\
$^{3}A_{2}$ & $n\rightarrow 3p$ & 9.61  & 9.49  & $-0.12$ & 9.61  & 9.50  & $-0.11$ \\
$^{3}A_{1}$ & $n\rightarrow 3s$ & 9.83  & 9.69  & $-0.14$ & 9.80  & 9.66  & $-0.14$ \\
\end{tabular}
\end{ruledtabular}
\end{table*}

\begin{table*}[t]
\caption{Same as Table~\ref{tab:h2o_excitations}, but for CH$_2$O.}
\label{tab:ch2o_excitations}
\begin{ruledtabular}
\begin{tabular}{llcccccc}
& & \multicolumn{3}{c}{Present work} & \multicolumn{3}{c}{Ref.~\cite{Loos2020dyn}} \\
State
& Character
& $\Omega^{\mathrm{static}}$
& $\Omega^{\mathrm{dyn}}$
& $\Delta$
& $\Omega^{\mathrm{static}}$
& $\Omega^{\mathrm{dyn}}$
& $\Delta$ \\
\hline
$^{1}A_{2}$ & $n\rightarrow\pi^{*}$   & 5.09 & 4.73 & $-0.36$ & 5.03 & 4.68 & $-0.35$ \\
$^{1}B_{2}$ & $n\rightarrow 3s$        & 7.84 & 7.82 & $-0.02$ & 7.87 & 7.85 & $-0.02$ \\
$^{1}B_{2}$ & $n\rightarrow 3p$        & 8.75 & 8.71 & $-0.04$ & 8.76 & 8.72 & $-0.04$ \\
$^{1}A_{1}$ & $n\rightarrow 3p$        & 8.84 & 8.82 & $-0.02$ & 8.85 & 8.84 & $-0.01$ \\
$^{1}A_{2}$ & $n\rightarrow 3p$        & 8.84 & 8.83 & $-0.01$ & 8.87 & 8.85 & $-0.02$ \\
$^{3}A_{2}$ & $n\rightarrow\pi^{*}$   & 4.36 & 3.94 & $-0.41$ & 4.28 & 3.87 & $-0.40$ \\
$^{3}A_{1}$ & $\pi\rightarrow\pi^{*}$ & 6.63 & 6.08 & $-0.55$ & 6.31 & 5.75 & $-0.56$ \\
$^{3}B_{2}$ & $n\rightarrow 3s$        & 7.57 & 7.51 & $-0.06$ & 7.60 & 7.56 & $-0.05$ \\
\end{tabular}
\end{ruledtabular}
\end{table*}

\begin{table*}[t]
\caption{Same as Table~\ref{tab:h2o_excitations}, but for C$_2$H$_2$.}
\label{tab:c2h2_excitations}
\begin{ruledtabular}
\begin{tabular}{llcccccc}
& & \multicolumn{3}{c}{Present work} & \multicolumn{3}{c}{Ref.~\cite{Loos2020dyn}} \\
State
& Character
& $\Omega^{\mathrm{static}}$
& $\Omega^{\mathrm{dyn}}$
& $\Delta$
& $\Omega^{\mathrm{static}}$
& $\Omega^{\mathrm{dyn}}$
& $\Delta$ \\
\hline
$^{1}\Sigma_{u}^{-}$ & $\pi\rightarrow\pi^{*}$ & 7.44 & 7.12 & $-0.33$ & 7.37 & 7.05 & $-0.32$ \\
$^{1}\Delta_{u}$     & $\pi\rightarrow\pi^{*}$ & 7.81 & 7.51 & $-0.30$ & 7.74 & 7.46 & $-0.29$ \\
$^{3}\Sigma_{u}^{+}$ & $\pi\rightarrow\pi^{*}$ & 6.15 & 5.65 & $-0.50$ & 5.83 & 5.32 & $-0.51$ \\
$^{3}\Delta_{u}$     & $\pi\rightarrow\pi^{*}$ & 6.81 & 6.40 & $-0.41$ & 6.64 & 6.23 & $-0.41$ \\
$^{3}\Sigma_{u}^{-}$ & $\pi\rightarrow\pi^{*}$ & 7.44 & 7.12 & $-0.33$ & 7.37 & 7.05 & $-0.32$ \\
\end{tabular}
\end{ruledtabular}
\end{table*}

Table~\ref{tab:h2o_excitations} shows that, for the diffuse Rydberg excitations of $\rm{H_2O}$, the present static TDA-BSE energies agree with the full static BSE results of Ref.~\cite{Loos2020dyn} to within $0.03$ eV. The Rydberg states of $\rm{CH_2O}$ similarly show close agreement. In contrast, the TDA generally produces a blueshift for the $n\rightarrow\pi^{*}$ and $\pi\rightarrow\pi^{*}$ valence excitations reported in Tables~\ref{tab:ch2o_excitations} and \ref{tab:c2h2_excitations}. This blueshift is on the order of $0.1$ eV for the singlet valence states, while differences of up to $0.3$ eV occur for some triplet states. These differences are consistent with neglecting resonant-antiresonant coupling in the present static TDA-BSE calculations.\cite{Ma2009}

Despite the differences between the static BSE treatments, the dynamical shifts obtained from the direct nonlinear solution agree with the renormalized first-order corrections of Loos and Blase to within $0.01$ eV for every state considered. The dynamical redshifts are smallest for the diffuse Rydberg excitations, ranging from $0.01$ to $0.15$ eV. Larger corrections of $0.36$--$0.41$ eV are obtained for the $n\rightarrow\pi^{*}$ states, while the $\pi\rightarrow\pi^{*}$ corrections range from $0.30$ to $0.55$ eV.

The close agreement in the dynamical shifts confirms the conclusion of Ref.~\cite{Loos2020dyn} that a renormalized first-order dynamical correction provides an accurate approximation for these low-lying molecular excitations. For the wide-gap molecules studied here, the low-lying excitation energies remain well separated from resonances of $\tilde{W}(\Omega)$. Consequently, the relevant BSE eigenvalue branches depend only weakly and smoothly on $\Omega$, explaining both the success of the first-order treatment and the rapid convergence of the direct nonlinear solution.

\section{Conclusions and Future Directions}

We have formulated a real-time approach for constructing the frequency-dependent screened interaction entering the dynamical BSE. A pair-density source perturbation is used to generate the retarded screened interaction through TDH propagation. The causal response is transformed to a time-ordered interaction, multiplied by phase factors representing non-interacting electron-hole pair propagation, and Fourier transformed to obtain $\Wt(\Omega)$. The resulting nonlinear TDA BSE is solved by diagonalizing the excitonic matrix on a grid of trial frequencies and locating roots numerically. For $\rm{H_2O}$, $\rm{CH_2O}$, and $\rm{C_2H_2}$, we find that the low-lying eigenvalue branches vary smoothly with $\Omega$, allowing the dynamical roots to be accurately determined from a small number of sampled frequencies.

The dynamical shifts obtained from the direct nonlinear solution agree with the renormalized first-order perturbative corrections of Loos and Blase to within $0.01$ eV for every state considered. Dynamical screening produces small redshifts for the diffuse Rydberg excitations and substantially larger corrections for the $n\rightarrow\pi^{*}$ and $\pi\rightarrow\pi^{*}$ valence excitations. These results validate the real-time implementation and confirm that a renormalized first-order correction is accurate for low-lying excitations of the wide-gap molecules studied here.

It is challenging to include the off-diagonal block of the dynamical BSE because of numerical divergences.\cite{Loos2020dyn} We will explore how to handle this issue in future work. It is also worth mentioning that using a $GW$ approximation to the single-particle self-energy can result in unphysical solutions to the BSE. This has been explored by others \cite{Romaniello_2009,Romaniello2009,Sangalli2011,Zhang2013,cunningham_many-body_2025}, and we will consider this in future work. 

The present deterministic implementation retains the formal $\mathcal{O}(N^6)$ scaling. This scaling can be reduced by combining density-fitting of the two-electron integrals with iterative eigensolvers that avoid full diagonalization of the BSE matrix.\cite{Bintrim2022,Wen2026} In addition to these cost-saving strategies, the real-time formulation can exploit stochastic sampling, opening a route toward dynamical BSE calculations of much larger molecular systems. We briefly discuss this stochastic extension below.

\subsection{Future Applications with Stochastic Methods}
The real-time formulation lends itself naturally to the stochastic methods we have developed in Refs.~\cite{Neuhauser2012,PhysRevLett.111.106402,neuhauser2014breaking,Rabani2015,vlcek_swift_2018,bradbury_bethesalpeter_2022,bradbury_optimized_2023,bradbury_deterministicfragmented-stochastic_2023,no_more_gap}. Rather than explicitly constructing all matrix elements of the frequency-dependent screened interaction in the orbital basis, only a small set of stochastic matrix elements could be calculated by using random linear combinations of the occupied and unoccupied orbitals. The corresponding polarization response can then be measured using stochastic real-time dynamics.\cite{neuhauser2014breaking,gao_sublinear_2015}

We illustrate briefly how the dynamical BSE could be brought to systems containing hundreds to potentially thousands of valence electrons. We have shown with Eq. (\ref{eq:wtilde_time}) that the matrix elements of $\tilde{W}(t)$ and $W(t)$ are related by
\begin{equation}
\begin{split}
\tilde{W}_{ab,ij}(t)
&=
W_{ab,ij}(t)e^{-\eta|t|}
\Biggl[
e^{-i(\varepsilon_b-\varepsilon_i+\Delta)t}\theta(t)
+
e^{i(\varepsilon_a-\varepsilon_j+\Delta)t}\theta(-t)
\Biggr].
\end{split}
\label{eq:stochastic_wtilde_time}
\end{equation}
This relation is particularly useful for stochastic sampling because the orbital-energy phase factors can be transferred to random orbitals restricted to the occupied and unoccupied subspaces. The corresponding sampled matrix element can be written as
\begin{equation}
\begin{split}
(
\bar{\gamma}\bar{\bar{\gamma}}
|
\tilde{W}(t)
|
\bar{\beta}\bar{\bar{\beta}})
&=
\theta(t)e^{-i\Delta t}e^{-\eta|t|}
(\bar{\gamma}\bar{\bar{\gamma}}(t)
|W(t)|\bar{\beta}(-t)\bar{\bar{\beta}})
\\
&\quad+
\theta(-t)e^{i\Delta t}e^{-\eta|t|}
(\bar{\gamma}(-t)\bar{\bar{\gamma}}|W(t)|\bar{\beta}\bar{\bar{\beta}}(t)).
\end{split}
\label{eq:random_wtilde}
\end{equation}

Here, $\bar{\gamma}$ and $\bar{\bar{\gamma}}$ denote random-sign linear combinations of MOs within the unoccupied subspace, while $\bar{\beta}$ and $\bar{\bar{\beta}}$ denote random linear combinations of MOs within the occupied subspace. For example, $\bar{\gamma}(r)=\sum_{a\in N_c}\pm\phi_a(r)$, with an analogous expression for a random occupied function. The time dependence of these random states is governed by the corresponding one-particle Hamiltonian, i.e., $|\bar{\gamma}(t)\rangle=e^{-iH_0t}|\bar{\gamma}(0)\rangle$,
where $H_0$ may be an accurate quasiparticle Hamiltonian obtained from an optimally tuned range-separated hybrid DFT reference.\cite{bradbury_deterministicfragmented-stochastic_2023} The scissor shift $\Delta$ present in Eqs. (\ref{eq:stochastic_wtilde_time})-(\ref{eq:random_wtilde}) can be calculated with the low-scaling stochastic $GW$ method.\cite{Allen2024,Thomas2026} The resulting time-dependent matrix elements can then be Fourier transformed to obtain the corresponding sampled matrix elements of $\tilde{W}(\Omega)$. 

\begin{acknowledgments}
We thank Chern Chuang, Nadine C. Bradbury and Dimitri Bazile for useful discussions. This work is supported by the National Science Foundation (NSF) under Grant No. CHE-2245253 and BSF Grant BSF-2024265.
\end{acknowledgments}

\section*{Declaration of Competing Interest}
The authors declare that they have no known competing financial interests or personal relationships that could have appeared to influence the work reported in this paper.

\section*{Data Availability Statement}
The data that support the findings of this study are available from the corresponding author upon reasonable request.

\appendix

\section{Derivation of the Dynamical Bethe-Salpeter Equation within the Tamm-Dancoff and Shindo Approximations}
\label{app:shindo_dbse_derivation}

We begin from the Schwinger functional-derivative technique for the two-particle correlation function.\cite{Strinati,Bechstedt2015} First, we introduce an external potential $U(1,2)$, and define
\begin{equation}
    L(1,2;1',2')
    =
    -\ii
    \left.
    \frac{\delta G(1,1')}
    {\delta U(2',2)}
    \right.
    \label{eq:app_l_functional}
\end{equation}
The inverse Green's function is then:
\begin{equation}
    G^{-1}(1,2)
    =
    G_0^{-1}(1,2)
    -
    U(1,2)
    -
    v_H(1)\delta(1,2)
    -
    \Sigma(1,2),
    \label{eq:app_g_inverse}
\end{equation}
where $v_H(1)=-\ii\int d2 v(1,2)G(2,2^+)$ is the Hartree potential and $\Sigma$ is the single-particle self-energy, encompassing all many-body exchange-correlation effects. 
By using the identity $\frac{\delta}{\delta U}(GG^{-1})=0$ and taking the external potential to zero, one arrives at the Bethe-Salpeter equation
\begin{align}
L(1,2;1',2')
={}&
L_0(1,2;1',2')
\nonumber\\
&+
\int
d3\,d4\,d5\,d6\,
L_0(1,4;1',3)
\Xi(3,5;4,6)
L(6,2;5,2'),
\label{eq:app_bse_realspace}
\end{align}
where
\begin{equation}
    L_0(1,2;1',2')
    =
    -\ii G(1,2')G(2,1')
    \label{eq:app_l0_realspace}
\end{equation}
and the BSE kernel is defined as:
\begin{equation}
    \Xi(3,5;4,6)
    =
    \ii
    \frac{
    \delta\left[
    v_H(3)\delta(3,4)+\Sigma(3,4)
    \right]
    }{
    \delta G(6,5)
    }.
    \label{eq:app_kernel_functional}
\end{equation}
Within the $GW$ approximation,
\begin{equation}
    \Sigma(3,4)
    =
    \ii G(3,4)W(3^+,4).
    \label{eq:app_gw_self_energy}
\end{equation}
Neglecting the functional derivative $\delta W/\delta G$ produces
\begin{equation}
\begin{split}
    \Xi(3,5;4,6)
    ={}&
    \delta(3,4)\delta(5,6)v(3,6)
    \\
    &-
    \delta(3,6)\delta(4,5)W(3,4).
\end{split}
\label{eq:app_gw_bse_kernel}
\end{equation}
The first term produces the repulsive bare exchange interaction, while the second produces the attractive screened direct interaction.

Following the frequency convention of Romaniello \textit{et al.},\cite{Romaniello2009} the equilibrium four-point function is written in terms of an external bosonic frequency $\Omega$ and two internal fermionic frequencies $\omega'$ and $\omega''$. Suppressing spatial and spin variables, the BSE becomes
\begin{align}
L(\Omega,\omega',\omega'')
={}&
L_0(\Omega,\omega',\omega'')
\nonumber\\
&-
\ii
G\!\left(\omega'+\frac{\Omega}{2}\right)
G\!\left(\omega'-\frac{\Omega}{2}\right)
v
\int
\frac{d\bar{\omega}}{2\pi}
L(\Omega,\bar{\omega},\omega'')
\nonumber\\
&+
\ii
G\!\left(\omega'+\frac{\Omega}{2}\right)
G\!\left(\omega'-\frac{\Omega}{2}\right)
\int
\frac{d\bar{\omega}}{2\pi}
W(\omega'-\bar{\omega})
L(\Omega,\bar{\omega},\omega''),
\label{eq:app_bse_three_freq}
\end{align}
with
\begin{equation}
\begin{split}
L_0(\Omega,\omega',\omega'')
={}&
-2\pi\ii\,
\delta(\omega'-\omega'')
\\
&\times
G\!\left(\omega'+\frac{\Omega}{2}\right)
G\!\left(\omega''-\frac{\Omega}{2}\right).
\end{split}
\label{eq:app_l0_three_freq}
\end{equation}

We now project onto the resonant electron-hole sector and assume real-valued orbitals. The projected correlation function is then:
\begin{align}
L_{ia,jb}(\Omega,\omega',\omega'')
={}&
\int
d x_1\,d x_2\,d x_{1'}\,d x_{2'}\,
\phi_i(x_1)\phi_a(x_{1'})
\nonumber\\
&\times
L(x_1,x_2;x_{1'},x_{2'};\Omega,\omega',\omega'')
\phi_j(x_{2'})\phi_b(x_2).
\label{eq:app_project_l}
\end{align}
The Tamm-Dancoff BSE in this basis is
\begin{align}
L_{ia,jb}(\Omega,\omega',\omega'')
={}&
L^0_{ia,jb}(\Omega,\omega',\omega'')
-\ii
G_a\!\left(\omega'+\frac{\Omega}{2}\right)
G_i\!\left(\omega'-\frac{\Omega}{2}\right)
\nonumber\\
&\times
\sum_{kc}\int\frac{d\omega_1}{2\pi}
\left[
\kappa(ia|kc)
-(ac|W(\omega'-\omega_1)|ik)
\right]
L_{kc,jb}(\Omega,\omega_1,\omega''),
\label{eq:app_projected_bse}
\end{align}
where $\kappa=2$ for singlet excitations and $\kappa=0$ for triplet excitations. The uncorrelated propagator is
\begin{equation}
\begin{split}
&L^0_{ia,jb}(\Omega,\omega',\omega'')
=
-2\pi\ii\,\delta(\omega'-\omega'')\delta_{ij}\delta_{ab}
\\
&\qquad\times
G_a\!\left(\omega'+\frac{\Omega}{2}\right)
G_i\!\left(\omega''-\frac{\Omega}{2}\right).
\end{split}
\label{eq:app_l0_projected}
\end{equation}
The quasiparticle Green's functions are
\begin{equation}
    G_a(z)
    =
    \frac{1}{z-\varepsilon_a+\ii\eta},
    \qquad
    G_i(z)
    =
    \frac{1}{z-\varepsilon_i-\ii\eta},
    \label{eq:app_particle_hole_g}
\end{equation}
where $\eta\to0^{+}$.
We define the correlation function with the internal fermionic frequencies integrated out,
\begin{equation}
    \tilde{L}_{ia,jb}(\Omega)
    =
    \int
    \frac{d\omega'\,d\omega''}{(2\pi)^2}
    L_{ia,jb}(\Omega,\omega',\omega'').
    \label{eq:app_ltilde_def}
\end{equation}
With frequency dependence in $W$, this integration does not close the BSE. We therefore adopt the Shindo-type approximation \cite{Strinati, Bechstedt2015,Zhang2023}
\begin{align}
L_{ia,jb}(\Omega,\omega',\omega'')
\approx{}&
\frac{
G_a\!\left(\omega'+\frac{\Omega}{2}\right)
G_i\!\left(\omega'-\frac{\Omega}{2}\right)
}{
N_{ai}(\Omega)
}
\tilde{L}_{ia,jb}(\Omega)
\nonumber\\
&\times
\frac{
G_b\!\left(\omega''+\frac{\Omega}{2}\right)
G_j\!\left(\omega''-\frac{\Omega}{2}\right)
}{
N_{bj}(\Omega)
},
\label{eq:app_shindo_ansatz}
\end{align}
where
\begin{equation}
\begin{split}
N_{ai}(\Omega)
&=
\int
\frac{d\bar{\omega}}{2\pi}
G_a\!\left(\bar{\omega}+\frac{\Omega}{2}\right)
G_i\!\left(\bar{\omega}-\frac{\Omega}{2}\right)
\\
&=
\frac{\ii}
{\Omega-(\varepsilon_a-\varepsilon_i)+\ii\eta}.
\end{split}
\label{eq:app_n_ai}
\end{equation}
The same expression, with $a,i$ replaced by $b,j$, defines $N_{bj}(\Omega)$. The normalization ensures that integration of Eq. (\ref{eq:app_shindo_ansatz}) over $\omega'$ and $\omega''$ recovers  $\tilde{L}_{ia,jb}(\Omega)$.

The non-interacting contribution is
\begin{equation}
    \tilde{L}^{0}_{ia,jb}(\Omega)
    =
    \frac{\delta_{ij}\delta_{ab}}
    {\Omega-(\varepsilon_a-\varepsilon_i)+\ii\eta},
    \label{eq:app_l0_tilde}
\end{equation}
while the bare exchange part is
\begin{equation}
    \tilde{L}^{x}_{ia,jb}(\Omega)
    =
    \frac{1}
    {\Omega-(\varepsilon_a-\varepsilon_i)+\ii\eta}
    \sum_{kc}
    \kappa(ia|kc)
    \tilde{L}_{kc,jb}(\Omega).
    \label{eq:app_lx_tilde}
\end{equation}
For the screened direct term, inserting the Shindo ansatz gives
\begin{align}
\tilde{L}^{d}_{ia,jb}(\Omega)
={}&
\ii
\sum_{kc}
\frac{
\tilde{L}_{kc,jb}(\Omega)
}{
N_{ck}(\Omega)
}
\int
\frac{d\omega'\,d\omega_1}{(2\pi)^2}
\nonumber\\
&\times
G_a\!\left(\omega'+\frac{\Omega}{2}\right)
G_i\!\left(\omega'-\frac{\Omega}{2}\right)
(ac|W(\omega'-\omega_1)|ik)
\nonumber\\
&\times
G_c\!\left(\omega_1+\frac{\Omega}{2}\right)
G_k\!\left(\omega_1-\frac{\Omega}{2}\right).
\label{eq:app_ld_before_integration}
\end{align}
Evaluating the two internal-frequency integrals by contour integration yields
\begin{equation}
    \tilde{L}^{d}_{ia,jb}(\Omega)
    =
    -
    \frac{1}
    {\Omega-(\varepsilon_a-\varepsilon_i)+\ii\eta}
    \sum_{kc}
    \tilde{W}_{ia,kc}(\Omega)
    \tilde{L}_{kc,jb}(\Omega),
    \label{eq:app_ld_tilde}
\end{equation}
where
\begin{align}
\tilde{W}_{ia,kc}(\Omega)
\equiv{}&
\tilde{W}_{ac,ik}(\Omega)
\nonumber\\
={}&
\frac{\ii}{2\pi}
\int_{-\infty}^{\infty}
d\omega\,
(ac|W(\omega)|ik)
\nonumber\\
&\times
\left[
\frac{1}
{\Omega-(\varepsilon_c-\varepsilon_i+\omega)+\ii\eta}
+
\frac{1}
{\Omega-(\varepsilon_a-\varepsilon_k-\omega)+\ii\eta}
\right].
\label{eq:app_wtilde}
\end{align}
Equivalently, for the matrix element coupling the electron-hole pairs $ia$ and $jb$,
\begin{align}
\tilde{W}_{ia,jb}(\Omega)
\equiv{}&
\tilde{W}_{ab,ij}(\Omega)
\nonumber\\
={}&
\frac{\ii}{2\pi}
\int_{-\infty}^{\infty}
d\omega\,
W_{ab,ij}(\omega)
\nonumber\\
&\times
\left[
\frac{1}
{\Omega-(\varepsilon_b-\varepsilon_i+\omega)+\ii\eta}
+
\frac{1}
{\Omega-(\varepsilon_a-\varepsilon_j-\omega)+\ii\eta}
\right].
\label{eq:app_wtilde_main_indices}
\end{align}
Combining the uncorrelated, bare exchange, and screened direct terms gives
\begin{align}
\tilde{L}_{ia,jb}(\Omega)
={}&
\frac{\delta_{ij}\delta_{ab}}
{\Omega-(\varepsilon_a-\varepsilon_i)+\ii\eta}
\nonumber\\
&+
\frac{1}
{\Omega-(\varepsilon_a-\varepsilon_i)+\ii\eta}
\sum_{kc}
\left[
\kappa(ia|kc)
-
\tilde{W}_{ia,kc}(\Omega)
\right]
\tilde{L}_{kc,jb}(\Omega).
\label{eq:app_ltilde_closed}
\end{align}
We define the frequency-dependent TDA BSE matrix as:
\begin{equation}
    \tilde{A}_{ia,jb}(\Omega)
    =
    (\varepsilon_a-\varepsilon_i)
    \delta_{ij}\delta_{ab}
    +
    \kappa(ia|jb)
    -
    \tilde{W}_{ia,jb}(\Omega).
    \label{eq:app_atilde_def}
\end{equation}
Equation~\eqref{eq:app_ltilde_closed} can then be written in matrix form:
\begin{equation}
    \tilde{L}(\Omega)
    =
    \left[
    (\Omega+\ii\eta)I
    -
    \tilde{A}(\Omega)
    \right]^{-1}.
    \label{eq:app_ltilde_resolvent}
\end{equation}

\bibliography{refs}

\begin{thebibliography}{57}%
\makeatletter
\providecommand \@ifxundefined [1]{%
 \@ifx{#1\undefined}
}%
\providecommand \@ifnum [1]{%
 \ifnum #1\expandafter \@firstoftwo
 \else \expandafter \@secondoftwo
 \fi
}%
\providecommand \@ifx [1]{%
 \ifx #1\expandafter \@firstoftwo
 \else \expandafter \@secondoftwo
 \fi
}%
\providecommand \natexlab [1]{#1}%
\providecommand \enquote  [1]{``#1''}%
\providecommand \bibnamefont  [1]{#1}%
\providecommand \bibfnamefont [1]{#1}%
\providecommand \citenamefont [1]{#1}%
\providecommand \href@noop [0]{\@secondoftwo}%
\providecommand \href [0]{\begingroup \@sanitize@url \@href}%
\providecommand \@href[1]{\@@startlink{#1}\@@href}%
\providecommand \@@href[1]{\endgroup#1\@@endlink}%
\providecommand \@sanitize@url [0]{\catcode `\\12\catcode `\$12\catcode `\&12\catcode `\#12\catcode `\^12\catcode `\_12\catcode `\%12\relax}%
\providecommand \@@startlink[1]{}%
\providecommand \@@endlink[0]{}%
\providecommand \url  [0]{\begingroup\@sanitize@url \@url }%
\providecommand \@url [1]{\endgroup\@href {#1}{\urlprefix }}%
\providecommand \urlprefix  [0]{URL }%
\providecommand \Eprint [0]{\href }%
\providecommand \doibase [0]{http://dx.doi.org/}%
\providecommand \selectlanguage [0]{\@gobble}%
\providecommand \bibinfo  [0]{\@secondoftwo}%
\providecommand \bibfield  [0]{\@secondoftwo}%
\providecommand \translation [1]{[#1]}%
\providecommand \BibitemOpen [0]{}%
\providecommand \bibitemStop [0]{}%
\providecommand \bibitemNoStop [0]{.\EOS\space}%
\providecommand \EOS [0]{\spacefactor3000\relax}%
\providecommand \BibitemShut  [1]{\csname bibitem#1\endcsname}%
\let\auto@bib@innerbib\@empty
\bibitem [{\citenamefont {Ullrich}(2025)}]{ullrich_snapshot_2025}%
  \BibitemOpen
  \bibfield  {author} {\bibinfo {author} {\bibfnamefont {C.~A.}\ \bibnamefont {Ullrich}},\ }\bibfield  {title} {\enquote {\bibinfo {title} {A snapshot of time-dependent density-functional theory},}\ }\href {\doibase 10.1063/5.0297117} {\bibfield  {journal} {\bibinfo  {journal} {APL Computational Physics}\ }\textbf {\bibinfo {volume} {1}},\ \bibinfo {pages} {020901} (\bibinfo {year} {2025})}\BibitemShut {NoStop}%
\bibitem [{\citenamefont {Maitra}\ \emph {et~al.}(2004)\citenamefont {Maitra}, \citenamefont {Zhang}, \citenamefont {Cave},\ and\ \citenamefont {Burke}}]{Maitra2004}%
  \BibitemOpen
  \bibfield  {author} {\bibinfo {author} {\bibfnamefont {N.~T.}\ \bibnamefont {Maitra}}, \bibinfo {author} {\bibfnamefont {F.}~\bibnamefont {Zhang}}, \bibinfo {author} {\bibfnamefont {R.~J.}\ \bibnamefont {Cave}}, \ and\ \bibinfo {author} {\bibfnamefont {K.}~\bibnamefont {Burke}},\ }\bibfield  {title} {\enquote {\bibinfo {title} {Double excitations within time-dependent density functional theory linear response},}\ }\href {\doibase 10.1063/1.1651060} {\bibfield  {journal} {\bibinfo  {journal} {The Journal of Chemical Physics}\ }\textbf {\bibinfo {volume} {120}},\ \bibinfo {pages} {5932--5937} (\bibinfo {year} {2004})}\BibitemShut {NoStop}%
\bibitem [{\citenamefont {Levine}\ \emph {et~al.}(2006)\citenamefont {Levine}, \citenamefont {Ko}, \citenamefont {Quenneville},\ and\ \citenamefont {MartÍnez}}]{levine_conical_2006}%
  \BibitemOpen
  \bibfield  {author} {\bibinfo {author} {\bibfnamefont {B.~G.}\ \bibnamefont {Levine}}, \bibinfo {author} {\bibfnamefont {C.}~\bibnamefont {Ko}}, \bibinfo {author} {\bibfnamefont {J.}~\bibnamefont {Quenneville}}, \ and\ \bibinfo {author} {\bibfnamefont {T.~J.}\ \bibnamefont {MartÍnez}},\ }\bibfield  {title} {\enquote {\bibinfo {title} {Conical intersections and double excitations in time-dependent density functional theory},}\ }\href {\doibase 10.1080/00268970500417762} {\bibfield  {journal} {\bibinfo  {journal} {Molecular Physics}\ }\textbf {\bibinfo {volume} {104}},\ \bibinfo {pages} {1039--1051} (\bibinfo {year} {2006})}\BibitemShut {NoStop}%
\bibitem [{\citenamefont {Romaniello}\ \emph {et~al.}(2009)\citenamefont {Romaniello}, \citenamefont {Sangalli}, \citenamefont {Berger}, \citenamefont {Sottile}, \citenamefont {Molinari}, \citenamefont {Reining},\ and\ \citenamefont {Onida}}]{Romaniello2009}%
  \BibitemOpen
  \bibfield  {author} {\bibinfo {author} {\bibfnamefont {P.}~\bibnamefont {Romaniello}}, \bibinfo {author} {\bibfnamefont {D.}~\bibnamefont {Sangalli}}, \bibinfo {author} {\bibfnamefont {J.~A.}\ \bibnamefont {Berger}}, \bibinfo {author} {\bibfnamefont {F.}~\bibnamefont {Sottile}}, \bibinfo {author} {\bibfnamefont {L.~G.}\ \bibnamefont {Molinari}}, \bibinfo {author} {\bibfnamefont {L.}~\bibnamefont {Reining}}, \ and\ \bibinfo {author} {\bibfnamefont {G.}~\bibnamefont {Onida}},\ }\bibfield  {title} {\enquote {\bibinfo {title} {Double excitations in finite systems},}\ }\href@noop {} {\bibfield  {journal} {\bibinfo  {journal} {The Journal of Chemical Physics}\ }\textbf {\bibinfo {volume} {130}} (\bibinfo {year} {2009})}\BibitemShut {NoStop}%
\bibitem [{\citenamefont {Sangalli}\ \emph {et~al.}(2011)\citenamefont {Sangalli}, \citenamefont {Romaniello}, \citenamefont {Onida},\ and\ \citenamefont {Marini}}]{Sangalli2011}%
  \BibitemOpen
  \bibfield  {author} {\bibinfo {author} {\bibfnamefont {D.}~\bibnamefont {Sangalli}}, \bibinfo {author} {\bibfnamefont {P.}~\bibnamefont {Romaniello}}, \bibinfo {author} {\bibfnamefont {G.}~\bibnamefont {Onida}}, \ and\ \bibinfo {author} {\bibfnamefont {A.}~\bibnamefont {Marini}},\ }\bibfield  {title} {\enquote {\bibinfo {title} {Double excitations in correlated systems: A many-body approach},}\ }\href@noop {} {\bibfield  {journal} {\bibinfo  {journal} {The Journal of Chemical Physics}\ }\textbf {\bibinfo {volume} {134}} (\bibinfo {year} {2011})}\BibitemShut {NoStop}%
\bibitem [{\citenamefont {Zhang}, \citenamefont {Steinmann},\ and\ \citenamefont {Yang}(2013)}]{Zhang2013}%
  \BibitemOpen
  \bibfield  {author} {\bibinfo {author} {\bibfnamefont {D.}~\bibnamefont {Zhang}}, \bibinfo {author} {\bibfnamefont {S.~N.}\ \bibnamefont {Steinmann}}, \ and\ \bibinfo {author} {\bibfnamefont {W.}~\bibnamefont {Yang}},\ }\bibfield  {title} {\enquote {\bibinfo {title} {Dynamical second-order {Bethe}-{Salpeter} equation kernel: {A} method for electronic excitation beyond the adiabatic approximation},}\ }\href {\doibase 10.1063/1.4824907} {\bibfield  {journal} {\bibinfo  {journal} {The Journal of Chemical Physics}\ }\textbf {\bibinfo {volume} {139}},\ \bibinfo {pages} {154109} (\bibinfo {year} {2013})}\BibitemShut {NoStop}%
\bibitem [{\citenamefont {Rebolini}\ and\ \citenamefont {Toulouse}(2016)}]{Rebolini2016}%
  \BibitemOpen
  \bibfield  {author} {\bibinfo {author} {\bibfnamefont {E.}~\bibnamefont {Rebolini}}\ and\ \bibinfo {author} {\bibfnamefont {J.}~\bibnamefont {Toulouse}},\ }\bibfield  {title} {\enquote {\bibinfo {title} {Range-separated time-dependent density-functional theory with a frequency-dependent second-order {Bethe}-{Salpeter} correlation kernel},}\ }\href {\doibase 10.1063/1.4943003} {\bibfield  {journal} {\bibinfo  {journal} {The Journal of Chemical Physics}\ }\textbf {\bibinfo {volume} {144}},\ \bibinfo {pages} {094107} (\bibinfo {year} {2016})}\BibitemShut {NoStop}%
\bibitem [{\citenamefont {Lacombe}\ and\ \citenamefont {Maitra}(2023)}]{Lacombe2023}%
  \BibitemOpen
  \bibfield  {author} {\bibinfo {author} {\bibfnamefont {L.}~\bibnamefont {Lacombe}}\ and\ \bibinfo {author} {\bibfnamefont {N.~T.}\ \bibnamefont {Maitra}},\ }\bibfield  {title} {\enquote {\bibinfo {title} {Non-adiabatic approximations in time-dependent density functional theory: progress and prospects},}\ }\href {http://dx.doi.org/10.1038/s41524-023-01061-0} {\bibfield  {journal} {\bibinfo  {journal} {npj Computational Materials}\ }\textbf {\bibinfo {volume} {9}} (\bibinfo {year} {2023})}\BibitemShut {NoStop}%
\bibitem [{\citenamefont {Hedin}(1965)}]{Hedin_1965}%
  \BibitemOpen
  \bibfield  {author} {\bibinfo {author} {\bibfnamefont {L.}~\bibnamefont {Hedin}},\ }\bibfield  {title} {\enquote {\bibinfo {title} {New method for calculating the {One-Particle Green's Function} with application to the electron-gas problem},}\ }\href {\doibase 10.1103/PhysRev.139.A796} {\bibfield  {journal} {\bibinfo  {journal} {Phys. Rev.}\ }\textbf {\bibinfo {volume} {139}},\ \bibinfo {pages} {A796--A823} (\bibinfo {year} {1965})}\BibitemShut {NoStop}%
\bibitem [{\citenamefont {Hybertsen}\ and\ \citenamefont {Louie}(1985)}]{Louie1985}%
  \BibitemOpen
  \bibfield  {author} {\bibinfo {author} {\bibfnamefont {M.~S.}\ \bibnamefont {Hybertsen}}\ and\ \bibinfo {author} {\bibfnamefont {S.~G.}\ \bibnamefont {Louie}},\ }\bibfield  {title} {\enquote {\bibinfo {title} {First-principles theory of quasiparticles: Calculation of band gaps in semiconductors and insulators},}\ }\href {\doibase 10.1103/physrevlett.55.1418} {\bibfield  {journal} {\bibinfo  {journal} {Physical Review Letters}\ }\textbf {\bibinfo {volume} {55}},\ \bibinfo {pages} {1418--1421} (\bibinfo {year} {1985})}\BibitemShut {NoStop}%
\bibitem [{\citenamefont {Strinati}(1988)}]{Strinati}%
  \BibitemOpen
  \bibfield  {author} {\bibinfo {author} {\bibfnamefont {G.}~\bibnamefont {Strinati}},\ }\bibfield  {title} {\enquote {\bibinfo {title} {Application of the {Green}’s functions method to the study of the optical properties of semiconductors},}\ }\href {\doibase 10.1007/BF02725962} {\bibfield  {journal} {\bibinfo  {journal} {La Rivista del Nuovo Cimento}\ }\textbf {\bibinfo {volume} {11}},\ \bibinfo {pages} {1--86} (\bibinfo {year} {1988})}\BibitemShut {NoStop}%
\bibitem [{\citenamefont {Blase}\ \emph {et~al.}(2020)\citenamefont {Blase}, \citenamefont {Duchemin}, \citenamefont {Jacquemin},\ and\ \citenamefont {Loos}}]{Blase2020}%
  \BibitemOpen
  \bibfield  {author} {\bibinfo {author} {\bibfnamefont {X.}~\bibnamefont {Blase}}, \bibinfo {author} {\bibfnamefont {I.}~\bibnamefont {Duchemin}}, \bibinfo {author} {\bibfnamefont {D.}~\bibnamefont {Jacquemin}}, \ and\ \bibinfo {author} {\bibfnamefont {P.-F.}\ \bibnamefont {Loos}},\ }\bibfield  {title} {\enquote {\bibinfo {title} {The {Bethe{\textendash}Salpeter Equation} formalism: From physics to chemistry},}\ }\href {\doibase 10.1021/acs.jpclett.0c01875} {\bibfield  {journal} {\bibinfo  {journal} {The Journal of Physical Chemistry Letters}\ }\textbf {\bibinfo {volume} {11}},\ \bibinfo {pages} {7371--7382} (\bibinfo {year} {2020})}\BibitemShut {NoStop}%
\bibitem [{\citenamefont {Rohlfing}\ and\ \citenamefont {Louie}(2000)}]{Louie_BSE_2000}%
  \BibitemOpen
  \bibfield  {author} {\bibinfo {author} {\bibfnamefont {M.}~\bibnamefont {Rohlfing}}\ and\ \bibinfo {author} {\bibfnamefont {S.~G.}\ \bibnamefont {Louie}},\ }\bibfield  {title} {\enquote {\bibinfo {title} {Electron-hole excitations and optical spectra from first principles},}\ }\href {\doibase 10.1103/PhysRevB.62.4927} {\bibfield  {journal} {\bibinfo  {journal} {Phys. Rev. B}\ }\textbf {\bibinfo {volume} {62}},\ \bibinfo {pages} {4927--4944} (\bibinfo {year} {2000})}\BibitemShut {NoStop}%
\bibitem [{\citenamefont {Onida}, \citenamefont {Reining},\ and\ \citenamefont {Rubio}(2002)}]{Onida_RMP_2002}%
  \BibitemOpen
  \bibfield  {author} {\bibinfo {author} {\bibfnamefont {G.}~\bibnamefont {Onida}}, \bibinfo {author} {\bibfnamefont {L.}~\bibnamefont {Reining}}, \ and\ \bibinfo {author} {\bibfnamefont {A.}~\bibnamefont {Rubio}},\ }\bibfield  {title} {\enquote {\bibinfo {title} {Electronic excitations: density-functional versus many-body {Green's}-function approaches},}\ }\href {\doibase 10.1103/RevModPhys.74.601} {\bibfield  {journal} {\bibinfo  {journal} {Rev. Mod. Phys.}\ }\textbf {\bibinfo {volume} {74}},\ \bibinfo {pages} {601--659} (\bibinfo {year} {2002})}\BibitemShut {NoStop}%
\bibitem [{\citenamefont {Jacquemin}, \citenamefont {Duchemin},\ and\ \citenamefont {Blase}(2017)}]{jacquemin_is_2017}%
  \BibitemOpen
  \bibfield  {author} {\bibinfo {author} {\bibfnamefont {D.}~\bibnamefont {Jacquemin}}, \bibinfo {author} {\bibfnamefont {I.}~\bibnamefont {Duchemin}}, \ and\ \bibinfo {author} {\bibfnamefont {X.}~\bibnamefont {Blase}},\ }\bibfield  {title} {\enquote {\bibinfo {title} {Is the {Bethe}–{Salpeter} {Formalism} {Accurate} for {Excitation} {Energies}? {Comparisons} with {TD}-{DFT}, {CASPT2}, and {EOM}-{CCSD}},}\ }\href {\doibase 10.1021/acs.jpclett.7b00381} {\bibfield  {journal} {\bibinfo  {journal} {The Journal of Physical Chemistry Letters}\ }\textbf {\bibinfo {volume} {8}},\ \bibinfo {pages} {1524--1529} (\bibinfo {year} {2017})}\BibitemShut {NoStop}%
\bibitem [{\citenamefont {Li}\ and\ \citenamefont {Olevano}(2022)}]{li_bethe-salpeter_2022}%
  \BibitemOpen
  \bibfield  {author} {\bibinfo {author} {\bibfnamefont {J.}~\bibnamefont {Li}}\ and\ \bibinfo {author} {\bibfnamefont {V.}~\bibnamefont {Olevano}},\ }\bibfield  {title} {\enquote {\bibinfo {title} {Bethe-{Salpeter} equation insights into the photo-absorption function and exciton structure of chlorophyll a and b in light-harvesting complex {II}},}\ }\href {\doibase 10.1016/j.jphotobiol.2022.112475} {\bibfield  {journal} {\bibinfo  {journal} {Journal of Photochemistry and Photobiology B: Biology}\ }\textbf {\bibinfo {volume} {232}},\ \bibinfo {pages} {112475} (\bibinfo {year} {2022})}\BibitemShut {NoStop}%
\bibitem [{\citenamefont {Rocca}\ \emph {et~al.}(2012)\citenamefont {Rocca}, \citenamefont {Ping}, \citenamefont {Gebauer},\ and\ \citenamefont {Galli}}]{Rocca2012-WEST}%
  \BibitemOpen
  \bibfield  {author} {\bibinfo {author} {\bibfnamefont {D.}~\bibnamefont {Rocca}}, \bibinfo {author} {\bibfnamefont {Y.}~\bibnamefont {Ping}}, \bibinfo {author} {\bibfnamefont {R.}~\bibnamefont {Gebauer}}, \ and\ \bibinfo {author} {\bibfnamefont {G.}~\bibnamefont {Galli}},\ }\bibfield  {title} {\enquote {\bibinfo {title} {Solution of the {Bethe-Salpeter} equation without empty electronic states: Application to the absorption spectra of bulk systems},}\ }\href {\doibase 10.1103/physrevb.85.045116} {\bibfield  {journal} {\bibinfo  {journal} {Physical Review B}\ }\textbf {\bibinfo {volume} {85}},\ \bibinfo {pages} {045116} (\bibinfo {year} {2012})}\BibitemShut {NoStop}%
\bibitem [{\citenamefont {Jacquemin}, \citenamefont {Duchemin},\ and\ \citenamefont {Blase}(2015)}]{Jacquemin2015}%
  \BibitemOpen
  \bibfield  {author} {\bibinfo {author} {\bibfnamefont {D.}~\bibnamefont {Jacquemin}}, \bibinfo {author} {\bibfnamefont {I.}~\bibnamefont {Duchemin}}, \ and\ \bibinfo {author} {\bibfnamefont {X.}~\bibnamefont {Blase}},\ }\bibfield  {title} {\enquote {\bibinfo {title} {Benchmarking the {Bethe-Salpeter} formalism on a standard organic molecular set},}\ }\href {\doibase 10.1021/acs.jctc.5b00304} {\bibfield  {journal} {\bibinfo  {journal} {Journal of Chemical Theory and Computation}\ }\textbf {\bibinfo {volume} {11}},\ \bibinfo {pages} {3290--3304} (\bibinfo {year} {2015})}\BibitemShut {NoStop}%
\bibitem [{\citenamefont {Gui}, \citenamefont {Holzer},\ and\ \citenamefont {Klopper}(2018)}]{Gui2018}%
  \BibitemOpen
  \bibfield  {author} {\bibinfo {author} {\bibfnamefont {X.}~\bibnamefont {Gui}}, \bibinfo {author} {\bibfnamefont {C.}~\bibnamefont {Holzer}}, \ and\ \bibinfo {author} {\bibfnamefont {W.}~\bibnamefont {Klopper}},\ }\bibfield  {title} {\enquote {\bibinfo {title} {Accuracy assessment of {$GW$} starting points for calculating molecular excitation energies using the {Bethe}{\textendash}{Salpeter} formalism},}\ }\href {\doibase 10.1021/acs.jctc.8b00014} {\bibfield  {journal} {\bibinfo  {journal} {Journal of Chemical Theory and Computation}\ }\textbf {\bibinfo {volume} {14}},\ \bibinfo {pages} {2127--2136} (\bibinfo {year} {2018})}\BibitemShut {NoStop}%
\bibitem [{\citenamefont {Förster}\ and\ \citenamefont {Visscher}(2022)}]{forster_quasiparticle_2022}%
  \BibitemOpen
  \bibfield  {author} {\bibinfo {author} {\bibfnamefont {A.}~\bibnamefont {Förster}}\ and\ \bibinfo {author} {\bibfnamefont {L.}~\bibnamefont {Visscher}},\ }\bibfield  {title} {\enquote {\bibinfo {title} {Quasiparticle self-consistent {$GW$-Bethe--Salpeter {Equation}} {Calculations} for {Large} {Chromophoric} {Systems}},}\ }\href {\doibase 10.1021/acs.jctc.2c00531} {\bibfield  {journal} {\bibinfo  {journal} {Journal of Chemical Theory and Computation}\ }\textbf {\bibinfo {volume} {18}},\ \bibinfo {pages} {6779--6793} (\bibinfo {year} {2022})}\BibitemShut {NoStop}%
\bibitem [{\citenamefont {Marini}\ and\ \citenamefont {Del~Sole}(2003)}]{marini_dynamical_2003}%
  \BibitemOpen
  \bibfield  {author} {\bibinfo {author} {\bibfnamefont {A.}~\bibnamefont {Marini}}\ and\ \bibinfo {author} {\bibfnamefont {R.}~\bibnamefont {Del~Sole}},\ }\bibfield  {title} {\enquote {\bibinfo {title} {Dynamical {Excitonic} {Effects} in {Metals} and {Semiconductors}},}\ }\href {\doibase 10.1103/PhysRevLett.91.176402} {\bibfield  {journal} {\bibinfo  {journal} {Physical Review Letters}\ }\textbf {\bibinfo {volume} {91}},\ \bibinfo {pages} {176402} (\bibinfo {year} {2003})}\BibitemShut {NoStop}%
\bibitem [{\citenamefont {Bechstedt}(2015)}]{Bechstedt2015}%
  \BibitemOpen
  \bibfield  {author} {\bibinfo {author} {\bibfnamefont {F.}~\bibnamefont {Bechstedt}},\ }\href {\doibase 10.1007/978-3-662-44593-8} {\emph {\bibinfo {title} {Many-Body Approach to Electronic Excitations: Concepts and Applications}}}\ (\bibinfo  {publisher} {Springer Berlin Heidelberg},\ \bibinfo {year} {2015})\BibitemShut {NoStop}%
\bibitem [{\citenamefont {Ma}, \citenamefont {Rohlfing},\ and\ \citenamefont {Molteni}(2009)}]{Ma2009}%
  \BibitemOpen
  \bibfield  {author} {\bibinfo {author} {\bibfnamefont {Y.}~\bibnamefont {Ma}}, \bibinfo {author} {\bibfnamefont {M.}~\bibnamefont {Rohlfing}}, \ and\ \bibinfo {author} {\bibfnamefont {C.}~\bibnamefont {Molteni}},\ }\bibfield  {title} {\enquote {\bibinfo {title} {Excited states of biological chromophores studied using many-body perturbation theory: {Effects} of resonant-antiresonant coupling and dynamical screening},}\ }\href {\doibase 10.1103/PhysRevB.80.241405} {\bibfield  {journal} {\bibinfo  {journal} {Physical Review B}\ }\textbf {\bibinfo {volume} {80}},\ \bibinfo {pages} {241405} (\bibinfo {year} {2009})}\BibitemShut {NoStop}%
\bibitem [{\citenamefont {Loos}\ and\ \citenamefont {Blase}(2020)}]{Loos2020dyn}%
  \BibitemOpen
  \bibfield  {author} {\bibinfo {author} {\bibfnamefont {P.-F.}\ \bibnamefont {Loos}}\ and\ \bibinfo {author} {\bibfnamefont {X.}~\bibnamefont {Blase}},\ }\bibfield  {title} {\enquote {\bibinfo {title} {Dynamical correction to the {Bethe-Salpeter} equation beyond the plasmon-pole approximation},}\ }\href {\doibase 10.1063/5.0023168} {\bibfield  {journal} {\bibinfo  {journal} {The Journal of Chemical Physics}\ }\textbf {\bibinfo {volume} {153}},\ \bibinfo {pages} {114120} (\bibinfo {year} {2020})}\BibitemShut {NoStop}%
\bibitem [{\citenamefont {Zhang}, \citenamefont {Leveillee},\ and\ \citenamefont {Schleife}(2023)}]{Zhang2023}%
  \BibitemOpen
  \bibfield  {author} {\bibinfo {author} {\bibfnamefont {X.}~\bibnamefont {Zhang}}, \bibinfo {author} {\bibfnamefont {J.~A.}\ \bibnamefont {Leveillee}}, \ and\ \bibinfo {author} {\bibfnamefont {A.}~\bibnamefont {Schleife}},\ }\bibfield  {title} {\enquote {\bibinfo {title} {Effect of dynamical screening in the {Bethe}-{Salpeter} framework: {Excitons} in crystalline naphthalene},}\ }\href {\doibase 10.1103/PhysRevB.107.235205} {\bibfield  {journal} {\bibinfo  {journal} {Physical Review B}\ }\textbf {\bibinfo {volume} {107}},\ \bibinfo {pages} {235205} (\bibinfo {year} {2023})}\BibitemShut {NoStop}%
\bibitem [{\citenamefont {Wen}, \citenamefont {Harsha},\ and\ \citenamefont {Zgid}(2026)}]{Wen2026}%
  \BibitemOpen
  \bibfield  {author} {\bibinfo {author} {\bibfnamefont {M.}~\bibnamefont {Wen}}, \bibinfo {author} {\bibfnamefont {G.}~\bibnamefont {Harsha}}, \ and\ \bibinfo {author} {\bibfnamefont {D.}~\bibnamefont {Zgid}},\ }\bibfield  {title} {\enquote {\bibinfo {title} {Dynamically {CorrectedBethe}–{Salpeter} {EquationSolver} for {Self}-{Consistent} \textit{{GW}} {Reference} on the {MatsubaraFrequency} {Axis}},}\ }\href@noop {} {\bibfield  {journal} {\bibinfo  {journal} {Journal of Chemical Theory and Computation}\ } (\bibinfo {year} {2026})}\BibitemShut {NoStop}%
\bibitem [{\citenamefont {Zhou}\ \emph {et~al.}(2026)\citenamefont {Zhou}, \citenamefont {Liu}, \citenamefont {Xu}, \citenamefont {Yao},\ and\ \citenamefont {Kanai}}]{Zhou2026}%
  \BibitemOpen
  \bibfield  {author} {\bibinfo {author} {\bibfnamefont {R.}~\bibnamefont {Zhou}}, \bibinfo {author} {\bibfnamefont {S.}~\bibnamefont {Liu}}, \bibinfo {author} {\bibfnamefont {J.}~\bibnamefont {Xu}}, \bibinfo {author} {\bibfnamefont {Y.}~\bibnamefont {Yao}}, \ and\ \bibinfo {author} {\bibfnamefont {Y.}~\bibnamefont {Kanai}},\ }\href@noop {} {\enquote {\bibinfo {title} {All-electron dynamical {Bethe-Salpeter Equation} for extended systems with atom-centered orbital basis set},}\ } (\bibinfo {year} {2026}),\ \Eprint {http://arxiv.org/abs/2606.08350} {arXiv:2606.08350 [physics.chem-ph]} \BibitemShut {NoStop}%
\bibitem [{\citenamefont {Bintrim}\ and\ \citenamefont {Berkelbach}(2022)}]{Bintrim2022}%
  \BibitemOpen
  \bibfield  {author} {\bibinfo {author} {\bibfnamefont {S.~J.}\ \bibnamefont {Bintrim}}\ and\ \bibinfo {author} {\bibfnamefont {T.~C.}\ \bibnamefont {Berkelbach}},\ }\bibfield  {title} {\enquote {\bibinfo {title} {Full-frequency dynamical {Bethe--Salpeter} equation without frequency and a study of double excitations},}\ }\href@noop {} {\bibfield  {journal} {\bibinfo  {journal} {The Journal of Chemical Physics}\ }\textbf {\bibinfo {volume} {156}} (\bibinfo {year} {2022})}\BibitemShut {NoStop}%
\bibitem [{\citenamefont {Kurzweil}\ and\ \citenamefont {Baer}(2008)}]{Kurzweil2008}%
  \BibitemOpen
  \bibfield  {author} {\bibinfo {author} {\bibfnamefont {Y.}~\bibnamefont {Kurzweil}}\ and\ \bibinfo {author} {\bibfnamefont {R.}~\bibnamefont {Baer}},\ }\bibfield  {title} {\enquote {\bibinfo {title} {Adapting approximate-memory potentials for time-dependent density functional theory},}\ }\href@noop {} {\bibfield  {journal} {\bibinfo  {journal} {Physical Review B}\ }\textbf {\bibinfo {volume} {77}} (\bibinfo {year} {2008})}\BibitemShut {NoStop}%
\bibitem [{\citenamefont {Daas}, \citenamefont {Crisostomo},\ and\ \citenamefont {Burke}(2026)}]{daas_prl_2026}%
  \BibitemOpen
  \bibfield  {author} {\bibinfo {author} {\bibfnamefont {K.~J.}\ \bibnamefont {Daas}}, \bibinfo {author} {\bibfnamefont {S.}~\bibnamefont {Crisostomo}}, \ and\ \bibinfo {author} {\bibfnamefont {K.}~\bibnamefont {Burke}},\ }\bibfield  {title} {\enquote {\bibinfo {title} {Ensemble time-dependent density functional theory},}\ }\href@noop {} {\bibfield  {journal} {\bibinfo  {journal} {Phys. Rev. Lett.}\ }\textbf {\bibinfo {volume} {137}},\ \bibinfo {pages} {028002} (\bibinfo {year} {2026})}\BibitemShut {NoStop}%
\bibitem [{\citenamefont {Casanova}\ and\ \citenamefont {Krylov}(2020)}]{casanova_spin-flip_2020}%
  \BibitemOpen
  \bibfield  {author} {\bibinfo {author} {\bibfnamefont {D.}~\bibnamefont {Casanova}}\ and\ \bibinfo {author} {\bibfnamefont {A.~I.}\ \bibnamefont {Krylov}},\ }\bibfield  {title} {\enquote {\bibinfo {title} {Spin-flip methods in quantum chemistry},}\ }\href {\doibase 10.1039/C9CP06507E} {\bibfield  {journal} {\bibinfo  {journal} {Physical Chemistry Chemical Physics}\ }\textbf {\bibinfo {volume} {22}},\ \bibinfo {pages} {4326--4342} (\bibinfo {year} {2020})}\BibitemShut {NoStop}%
\bibitem [{\citenamefont {Park}\ \emph {et~al.}(2021)\citenamefont {Park}, \citenamefont {Shen}, \citenamefont {Lee}, \citenamefont {Piecuch}, \citenamefont {Filatov},\ and\ \citenamefont {Choi}}]{Park2021}%
  \BibitemOpen
  \bibfield  {author} {\bibinfo {author} {\bibfnamefont {W.}~\bibnamefont {Park}}, \bibinfo {author} {\bibfnamefont {J.}~\bibnamefont {Shen}}, \bibinfo {author} {\bibfnamefont {S.}~\bibnamefont {Lee}}, \bibinfo {author} {\bibfnamefont {P.}~\bibnamefont {Piecuch}}, \bibinfo {author} {\bibfnamefont {M.}~\bibnamefont {Filatov}}, \ and\ \bibinfo {author} {\bibfnamefont {C.~H.}\ \bibnamefont {Choi}},\ }\bibfield  {title} {\enquote {\bibinfo {title} {Internal conversion between bright ({$1^1B_u^+$}) and dark ({$2^1A_g^-$}) states in s-trans-butadiene and s-trans-hexatriene},}\ }\href {\doibase 10.1021/acs.jpclett.1c02707} {\bibfield  {journal} {\bibinfo  {journal} {The Journal of Physical Chemistry Letters}\ }\textbf {\bibinfo {volume} {12}},\ \bibinfo {pages} {9720--9729} (\bibinfo {year} {2021})}\BibitemShut {NoStop}%
\bibitem [{\citenamefont {Dar}\ and\ \citenamefont {Maitra}(2025)}]{Dar2025}%
  \BibitemOpen
  \bibfield  {author} {\bibinfo {author} {\bibfnamefont {D.~B.}\ \bibnamefont {Dar}}\ and\ \bibinfo {author} {\bibfnamefont {N.~T.}\ \bibnamefont {Maitra}},\ }\bibfield  {title} {\enquote {\bibinfo {title} {Capturing the elusive curve-crossing in low-lying states of butadiene with {Dressed TDDFT}},}\ }\href {\doibase 10.1021/acs.jpclett.4c03167} {\bibfield  {journal} {\bibinfo  {journal} {The Journal of Physical Chemistry Letters}\ }\textbf {\bibinfo {volume} {16}},\ \bibinfo {pages} {703--709} (\bibinfo {year} {2025})}\BibitemShut {NoStop}%
\bibitem [{\citenamefont {Authier}\ and\ \citenamefont {Loos}(2020)}]{Authier2020}%
  \BibitemOpen
  \bibfield  {author} {\bibinfo {author} {\bibfnamefont {J.}~\bibnamefont {Authier}}\ and\ \bibinfo {author} {\bibfnamefont {P.-F.}\ \bibnamefont {Loos}},\ }\bibfield  {title} {\enquote {\bibinfo {title} {Dynamical kernels for optical excitations},}\ }\href {\doibase 10.1063/5.0028040} {\bibfield  {journal} {\bibinfo  {journal} {The Journal of Chemical Physics}\ }\textbf {\bibinfo {volume} {153}},\ \bibinfo {pages} {184105} (\bibinfo {year} {2020})}\BibitemShut {NoStop}%
\bibitem [{\citenamefont {Loos}\ and\ \citenamefont {Romaniello}(2022)}]{Loos2022}%
  \BibitemOpen
  \bibfield  {author} {\bibinfo {author} {\bibfnamefont {P.-F.}\ \bibnamefont {Loos}}\ and\ \bibinfo {author} {\bibfnamefont {P.}~\bibnamefont {Romaniello}},\ }\bibfield  {title} {\enquote {\bibinfo {title} {Static and dynamic {Bethe}–{Salpeter} equations in the \textit{{T}} -matrix approximation},}\ }\href {\doibase 10.1063/5.0088364} {\bibfield  {journal} {\bibinfo  {journal} {The Journal of Chemical Physics}\ }\textbf {\bibinfo {volume} {156}},\ \bibinfo {pages} {164101} (\bibinfo {year} {2022})}\BibitemShut {NoStop}%
\bibitem [{\citenamefont {Wilson}, \citenamefont {Gygi},\ and\ \citenamefont {Galli}(2008)}]{PhysRevB.78.113303}%
  \BibitemOpen
  \bibfield  {author} {\bibinfo {author} {\bibfnamefont {H.~F.}\ \bibnamefont {Wilson}}, \bibinfo {author} {\bibfnamefont {F.~m.~c.}\ \bibnamefont {Gygi}}, \ and\ \bibinfo {author} {\bibfnamefont {G.}~\bibnamefont {Galli}},\ }\bibfield  {title} {\enquote {\bibinfo {title} {Efficient iterative method for calculations of dielectric matrices},}\ }\href {\doibase 10.1103/PhysRevB.78.113303} {\bibfield  {journal} {\bibinfo  {journal} {Phys. Rev. B}\ }\textbf {\bibinfo {volume} {78}},\ \bibinfo {pages} {113303} (\bibinfo {year} {2008})}\BibitemShut {NoStop}%
\bibitem [{\citenamefont {Ren}\ \emph {et~al.}(2012)\citenamefont {Ren}, \citenamefont {Rinke}, \citenamefont {Blum}, \citenamefont {Wieferink}, \citenamefont {Tkatchenko}, \citenamefont {Sanfilippo}, \citenamefont {Reuter},\ and\ \citenamefont {Scheffler}}]{ren_resolution--identity_2012}%
  \BibitemOpen
  \bibfield  {author} {\bibinfo {author} {\bibfnamefont {X.}~\bibnamefont {Ren}}, \bibinfo {author} {\bibfnamefont {P.}~\bibnamefont {Rinke}}, \bibinfo {author} {\bibfnamefont {V.}~\bibnamefont {Blum}}, \bibinfo {author} {\bibfnamefont {J.}~\bibnamefont {Wieferink}}, \bibinfo {author} {\bibfnamefont {A.}~\bibnamefont {Tkatchenko}}, \bibinfo {author} {\bibfnamefont {A.}~\bibnamefont {Sanfilippo}}, \bibinfo {author} {\bibfnamefont {K.}~\bibnamefont {Reuter}}, \ and\ \bibinfo {author} {\bibfnamefont {M.}~\bibnamefont {Scheffler}},\ }\bibfield  {title} {\enquote {\bibinfo {title} {Resolution-of-identity approach to {Hartree}–{Fock}, hybrid density functionals, {RPA}, {MP2} and \textit{{GW}} with numeric atom-centered orbital basis functions},}\ }\href@noop {} {\bibfield  {journal} {\bibinfo  {journal} {New Journal of Physics}\ }\textbf {\bibinfo {volume} {14}},\ \bibinfo {pages} {053020} (\bibinfo {year} {2012})}\BibitemShut {NoStop}%
\bibitem [{\citenamefont {Neuhauser}\ \emph {et~al.}(2014)\citenamefont {Neuhauser}, \citenamefont {Gao}, \citenamefont {Arntsen}, \citenamefont {Karshenas}, \citenamefont {Rabani},\ and\ \citenamefont {Baer}}]{neuhauser2014breaking}%
  \BibitemOpen
  \bibfield  {author} {\bibinfo {author} {\bibfnamefont {D.}~\bibnamefont {Neuhauser}}, \bibinfo {author} {\bibfnamefont {Y.}~\bibnamefont {Gao}}, \bibinfo {author} {\bibfnamefont {C.}~\bibnamefont {Arntsen}}, \bibinfo {author} {\bibfnamefont {C.}~\bibnamefont {Karshenas}}, \bibinfo {author} {\bibfnamefont {E.}~\bibnamefont {Rabani}}, \ and\ \bibinfo {author} {\bibfnamefont {R.}~\bibnamefont {Baer}},\ }\bibfield  {title} {\enquote {\bibinfo {title} {Breaking the theoretical scaling limit for predicting quasiparticle energies: The stochastic ${GW}$ approach},}\ }\href@noop {} {\bibfield  {journal} {\bibinfo  {journal} {Phys. Rev. Lett.}\ }\textbf {\bibinfo {volume} {113}},\ \bibinfo {pages} {076402} (\bibinfo {year} {2014})}\BibitemShut {NoStop}%
\bibitem [{\citenamefont {Duchemin}\ and\ \citenamefont {Blase}(2021)}]{duchemin_cubic-scaling_2021}%
  \BibitemOpen
  \bibfield  {author} {\bibinfo {author} {\bibfnamefont {I.}~\bibnamefont {Duchemin}}\ and\ \bibinfo {author} {\bibfnamefont {X.}~\bibnamefont {Blase}},\ }\bibfield  {title} {\enquote {\bibinfo {title} {Cubic-{Scaling} {All}-{Electron} \textit{{GW}} {Calculations} with a {Separable} {Density}-{Fitting} {Space}–{Time} {Approach}},}\ }\href@noop {} {\bibfield  {journal} {\bibinfo  {journal} {Journal of Chemical Theory and Computation}\ }\textbf {\bibinfo {volume} {17}},\ \bibinfo {pages} {2383--2393} (\bibinfo {year} {2021})}\BibitemShut {NoStop}%
\bibitem [{\citenamefont {Yeh}\ \emph {et~al.}(2022)\citenamefont {Yeh}, \citenamefont {Iskakov}, \citenamefont {Zgid},\ and\ \citenamefont {Gull}}]{yeh_fully_2022}%
  \BibitemOpen
  \bibfield  {author} {\bibinfo {author} {\bibfnamefont {C.-N.}\ \bibnamefont {Yeh}}, \bibinfo {author} {\bibfnamefont {S.}~\bibnamefont {Iskakov}}, \bibinfo {author} {\bibfnamefont {D.}~\bibnamefont {Zgid}}, \ and\ \bibinfo {author} {\bibfnamefont {E.}~\bibnamefont {Gull}},\ }\bibfield  {title} {\enquote {\bibinfo {title} {Fully self-consistent finite-temperature {G} {W} in {Gaussian} {Bloch} orbitals for solids},}\ }\href@noop {} {\bibfield  {journal} {\bibinfo  {journal} {Physical Review B}\ }\textbf {\bibinfo {volume} {106}},\ \bibinfo {pages} {235104} (\bibinfo {year} {2022})}\BibitemShut {NoStop}%
\bibitem [{\citenamefont {Rabani}, \citenamefont {Baer},\ and\ \citenamefont {Neuhauser}(2015)}]{Rabani2015}%
  \BibitemOpen
  \bibfield  {author} {\bibinfo {author} {\bibfnamefont {E.}~\bibnamefont {Rabani}}, \bibinfo {author} {\bibfnamefont {R.}~\bibnamefont {Baer}}, \ and\ \bibinfo {author} {\bibfnamefont {D.}~\bibnamefont {Neuhauser}},\ }\bibfield  {title} {\enquote {\bibinfo {title} {Time-dependent stochastic {Bethe-Salpeter} approach},}\ }\href {\doibase 10.1103/physrevb.91.235302} {\bibfield  {journal} {\bibinfo  {journal} {Physical Review B}\ }\textbf {\bibinfo {volume} {91}},\ \bibinfo {pages} {235302} (\bibinfo {year} {2015})}\BibitemShut {NoStop}%
\bibitem [{\citenamefont {Bradbury}\ \emph {et~al.}(2023{\natexlab{a}})\citenamefont {Bradbury}, \citenamefont {Allen}, \citenamefont {Nguyen}, \citenamefont {Ibrahim},\ and\ \citenamefont {Neuhauser}}]{bradbury_optimized_2023}%
  \BibitemOpen
  \bibfield  {author} {\bibinfo {author} {\bibfnamefont {N.~C.}\ \bibnamefont {Bradbury}}, \bibinfo {author} {\bibfnamefont {T.}~\bibnamefont {Allen}}, \bibinfo {author} {\bibfnamefont {M.}~\bibnamefont {Nguyen}}, \bibinfo {author} {\bibfnamefont {K.~Z.}\ \bibnamefont {Ibrahim}}, \ and\ \bibinfo {author} {\bibfnamefont {D.}~\bibnamefont {Neuhauser}},\ }\bibfield  {title} {\enquote {\bibinfo {title} {Optimized attenuated interaction: {Enabling} stochastic {Bethe}–{Salpeter} spectra for large systems},}\ }\href {\doibase 10.1063/5.0146555} {\bibfield  {journal} {\bibinfo  {journal} {The Journal of Chemical Physics}\ }\textbf {\bibinfo {volume} {158}},\ \bibinfo {pages} {154104} (\bibinfo {year} {2023}{\natexlab{a}})}\BibitemShut {NoStop}%
\bibitem [{\citenamefont {Hillenbrand}, \citenamefont {Li},\ and\ \citenamefont {Zhu}(2025)}]{hillenbrand_energy-specific_2025}%
  \BibitemOpen
  \bibfield  {author} {\bibinfo {author} {\bibfnamefont {C.}~\bibnamefont {Hillenbrand}}, \bibinfo {author} {\bibfnamefont {J.}~\bibnamefont {Li}}, \ and\ \bibinfo {author} {\bibfnamefont {T.}~\bibnamefont {Zhu}},\ }\bibfield  {title} {\enquote {\bibinfo {title} {Energy-specific {Bethe}–{Salpeter} equation implementation for efficient optical spectrum calculations},}\ }\href {\doibase 10.1063/5.0260895} {\bibfield  {journal} {\bibinfo  {journal} {The Journal of Chemical Physics}\ }\textbf {\bibinfo {volume} {162}},\ \bibinfo {pages} {174117} (\bibinfo {year} {2025})}\BibitemShut {NoStop}%
\bibitem [{\citenamefont {Vlček}\ \emph {et~al.}(2018)\citenamefont {Vlček}, \citenamefont {Li}, \citenamefont {Baer}, \citenamefont {Rabani},\ and\ \citenamefont {Neuhauser}}]{vlcek_swift_2018}%
  \BibitemOpen
  \bibfield  {author} {\bibinfo {author} {\bibfnamefont {V.}~\bibnamefont {Vlček}}, \bibinfo {author} {\bibfnamefont {W.}~\bibnamefont {Li}}, \bibinfo {author} {\bibfnamefont {R.}~\bibnamefont {Baer}}, \bibinfo {author} {\bibfnamefont {E.}~\bibnamefont {Rabani}}, \ and\ \bibinfo {author} {\bibfnamefont {D.}~\bibnamefont {Neuhauser}},\ }\bibfield  {title} {\enquote {\bibinfo {title} {Swift {$GW$} beyond 10,000 electrons using sparse stochastic compression},}\ }\href {\doibase 10.1103/PhysRevB.98.075107} {\bibfield  {journal} {\bibinfo  {journal} {Physical Review B}\ }\textbf {\bibinfo {volume} {98}},\ \bibinfo {pages} {075107} (\bibinfo {year} {2018})}\BibitemShut {NoStop}%
\bibitem [{\citenamefont {Bradbury}\ \emph {et~al.}(2022)\citenamefont {Bradbury}, \citenamefont {Nguyen}, \citenamefont {Caram},\ and\ \citenamefont {Neuhauser}}]{bradbury_bethesalpeter_2022}%
  \BibitemOpen
  \bibfield  {author} {\bibinfo {author} {\bibfnamefont {N.~C.}\ \bibnamefont {Bradbury}}, \bibinfo {author} {\bibfnamefont {M.}~\bibnamefont {Nguyen}}, \bibinfo {author} {\bibfnamefont {J.~R.}\ \bibnamefont {Caram}}, \ and\ \bibinfo {author} {\bibfnamefont {D.}~\bibnamefont {Neuhauser}},\ }\bibfield  {title} {\enquote {\bibinfo {title} {Bethe–{Salpeter} equation spectra for very large systems},}\ }\href {\doibase 10.1063/5.0100213} {\bibfield  {journal} {\bibinfo  {journal} {The Journal of Chemical Physics}\ }\textbf {\bibinfo {volume} {157}},\ \bibinfo {pages} {031104} (\bibinfo {year} {2022})}\BibitemShut {NoStop}%
\bibitem [{\citenamefont {Bradbury}\ \emph {et~al.}(2024)\citenamefont {Bradbury}, \citenamefont {Li}, \citenamefont {Allen}, \citenamefont {Caram},\ and\ \citenamefont {Neuhauser}}]{no_more_gap}%
  \BibitemOpen
  \bibfield  {author} {\bibinfo {author} {\bibfnamefont {N.~C.}\ \bibnamefont {Bradbury}}, \bibinfo {author} {\bibfnamefont {B.~Y.}\ \bibnamefont {Li}}, \bibinfo {author} {\bibfnamefont {T.}~\bibnamefont {Allen}}, \bibinfo {author} {\bibfnamefont {J.~R.}\ \bibnamefont {Caram}}, \ and\ \bibinfo {author} {\bibfnamefont {D.}~\bibnamefont {Neuhauser}},\ }\bibfield  {title} {\enquote {\bibinfo {title} {No more gap-shifting: Stochastic many-body-theory based {TDHF} for accurate theory of polymethine cyanine dyes},}\ }\href {\doibase 10.1063/5.0223783} {\bibfield  {journal} {\bibinfo  {journal} {The Journal of Chemical Physics}\ }\textbf {\bibinfo {volume} {161}},\ \bibinfo {pages} {141101} (\bibinfo {year} {2024})}\BibitemShut {NoStop}%
\bibitem [{\citenamefont {Allen}\ \emph {et~al.}(2026)\citenamefont {Allen}, \citenamefont {Li}, \citenamefont {Bradbury},\ and\ \citenamefont {Neuhauser}}]{Allen2026}%
  \BibitemOpen
  \bibfield  {author} {\bibinfo {author} {\bibfnamefont {T.}~\bibnamefont {Allen}}, \bibinfo {author} {\bibfnamefont {B.~Y.}\ \bibnamefont {Li}}, \bibinfo {author} {\bibfnamefont {N.~C.}\ \bibnamefont {Bradbury}}, \ and\ \bibinfo {author} {\bibfnamefont {D.}~\bibnamefont {Neuhauser}},\ }\bibfield  {title} {\enquote {\bibinfo {title} {Environment-{Induced} {Exciton} {Renormalization} in the {Photosystem} {II} {Reaction} {Center}},}\ }\href {\doibase 10.1021/acs.jctc.6c00440} {\bibfield  {journal} {\bibinfo  {journal} {Journal of Chemical Theory and Computation}\ ,\ \bibinfo {pages} {acs.jctc.6c00440}} (\bibinfo {year} {2026})}\BibitemShut {NoStop}%
\bibitem [{\citenamefont {Sun}\ \emph {et~al.}(2020)\citenamefont {Sun}, \citenamefont {Zhang}, \citenamefont {Banerjee}, \citenamefont {Bao}, \citenamefont {Barbry}, \citenamefont {Blunt}, \citenamefont {Bogdanov}, \citenamefont {Booth}, \citenamefont {Chen}, \citenamefont {Cui}, \citenamefont {Eriksen}, \citenamefont {Gao}, \citenamefont {Guo}, \citenamefont {Hermann}, \citenamefont {Hermes}, \citenamefont {Koh}, \citenamefont {Koval}, \citenamefont {Lehtola}, \citenamefont {Li}, \citenamefont {Liu}, \citenamefont {Mardirossian}, \citenamefont {McClain}, \citenamefont {Motta}, \citenamefont {Mussard}, \citenamefont {Pham}, \citenamefont {Pulkin}, \citenamefont {Purwanto}, \citenamefont {Robinson}, \citenamefont {Ronca}, \citenamefont {Sayfutyarova}, \citenamefont {Scheurer}, \citenamefont {Schurkus}, \citenamefont {Smith}, \citenamefont {Sun}, \citenamefont {Sun}, \citenamefont {Upadhyay}, \citenamefont {Wagner}, \citenamefont {Wang}, \citenamefont {White}, \citenamefont {Whitfield}, \citenamefont
  {Williamson}, \citenamefont {Wouters}, \citenamefont {Yang}, \citenamefont {Yu}, \citenamefont {Zhu}, \citenamefont {Berkelbach}, \citenamefont {Sharma}, \citenamefont {Sokolov},\ and\ \citenamefont {Chan}}]{sun_recent_2020}%
  \BibitemOpen
  \bibfield  {author} {\bibinfo {author} {\bibfnamefont {Q.}~\bibnamefont {Sun}}, \bibinfo {author} {\bibfnamefont {X.}~\bibnamefont {Zhang}}, \bibinfo {author} {\bibfnamefont {S.}~\bibnamefont {Banerjee}}, \bibinfo {author} {\bibfnamefont {P.}~\bibnamefont {Bao}}, \bibinfo {author} {\bibfnamefont {M.}~\bibnamefont {Barbry}}, \bibinfo {author} {\bibfnamefont {N.~S.}\ \bibnamefont {Blunt}}, \bibinfo {author} {\bibfnamefont {N.~A.}\ \bibnamefont {Bogdanov}}, \bibinfo {author} {\bibfnamefont {G.~H.}\ \bibnamefont {Booth}}, \bibinfo {author} {\bibfnamefont {J.}~\bibnamefont {Chen}}, \bibinfo {author} {\bibfnamefont {Z.-H.}\ \bibnamefont {Cui}}, \bibinfo {author} {\bibfnamefont {J.~J.}\ \bibnamefont {Eriksen}}, \bibinfo {author} {\bibfnamefont {Y.}~\bibnamefont {Gao}}, \bibinfo {author} {\bibfnamefont {S.}~\bibnamefont {Guo}}, \bibinfo {author} {\bibfnamefont {J.}~\bibnamefont {Hermann}}, \bibinfo {author} {\bibfnamefont {M.~R.}\ \bibnamefont {Hermes}}, \bibinfo {author} {\bibfnamefont {K.}~\bibnamefont {Koh}},
  \bibinfo {author} {\bibfnamefont {P.}~\bibnamefont {Koval}}, \bibinfo {author} {\bibfnamefont {S.}~\bibnamefont {Lehtola}}, \bibinfo {author} {\bibfnamefont {Z.}~\bibnamefont {Li}}, \bibinfo {author} {\bibfnamefont {J.}~\bibnamefont {Liu}}, \bibinfo {author} {\bibfnamefont {N.}~\bibnamefont {Mardirossian}}, \bibinfo {author} {\bibfnamefont {J.~D.}\ \bibnamefont {McClain}}, \bibinfo {author} {\bibfnamefont {M.}~\bibnamefont {Motta}}, \bibinfo {author} {\bibfnamefont {B.}~\bibnamefont {Mussard}}, \bibinfo {author} {\bibfnamefont {H.~Q.}\ \bibnamefont {Pham}}, \bibinfo {author} {\bibfnamefont {A.}~\bibnamefont {Pulkin}}, \bibinfo {author} {\bibfnamefont {W.}~\bibnamefont {Purwanto}}, \bibinfo {author} {\bibfnamefont {P.~J.}\ \bibnamefont {Robinson}}, \bibinfo {author} {\bibfnamefont {E.}~\bibnamefont {Ronca}}, \bibinfo {author} {\bibfnamefont {E.~R.}\ \bibnamefont {Sayfutyarova}}, \bibinfo {author} {\bibfnamefont {M.}~\bibnamefont {Scheurer}}, \bibinfo {author} {\bibfnamefont {H.~F.}\ \bibnamefont {Schurkus}},
  \bibinfo {author} {\bibfnamefont {J.~E.~T.}\ \bibnamefont {Smith}}, \bibinfo {author} {\bibfnamefont {C.}~\bibnamefont {Sun}}, \bibinfo {author} {\bibfnamefont {S.-N.}\ \bibnamefont {Sun}}, \bibinfo {author} {\bibfnamefont {S.}~\bibnamefont {Upadhyay}}, \bibinfo {author} {\bibfnamefont {L.~K.}\ \bibnamefont {Wagner}}, \bibinfo {author} {\bibfnamefont {X.}~\bibnamefont {Wang}}, \bibinfo {author} {\bibfnamefont {A.}~\bibnamefont {White}}, \bibinfo {author} {\bibfnamefont {J.~D.}\ \bibnamefont {Whitfield}}, \bibinfo {author} {\bibfnamefont {M.~J.}\ \bibnamefont {Williamson}}, \bibinfo {author} {\bibfnamefont {S.}~\bibnamefont {Wouters}}, \bibinfo {author} {\bibfnamefont {J.}~\bibnamefont {Yang}}, \bibinfo {author} {\bibfnamefont {J.~M.}\ \bibnamefont {Yu}}, \bibinfo {author} {\bibfnamefont {T.}~\bibnamefont {Zhu}}, \bibinfo {author} {\bibfnamefont {T.~C.}\ \bibnamefont {Berkelbach}}, \bibinfo {author} {\bibfnamefont {S.}~\bibnamefont {Sharma}}, \bibinfo {author} {\bibfnamefont {A.~Y.}\ \bibnamefont {Sokolov}},
  \ and\ \bibinfo {author} {\bibfnamefont {G.~K.-L.}\ \bibnamefont {Chan}},\ }\bibfield  {title} {\enquote {\bibinfo {title} {Recent developments in the {P} {\textless}span style="font-variant:small-caps;"{\textgreater}y{\textless}/span{\textgreater} {SCF} program package},}\ }\href {\doibase 10.1063/5.0006074} {\bibfield  {journal} {\bibinfo  {journal} {The Journal of Chemical Physics}\ }\textbf {\bibinfo {volume} {153}},\ \bibinfo {pages} {024109} (\bibinfo {year} {2020})}\BibitemShut {NoStop}%
\bibitem [{\citenamefont {Cunningham}(2025)}]{cunningham_many-body_2025}%
  \BibitemOpen
  \bibfield  {author} {\bibinfo {author} {\bibfnamefont {B.}~\bibnamefont {Cunningham}},\ }\bibfield  {title} {\enquote {\bibinfo {title} {Many-body vertex effects: {Time}-dependent interaction kernel with correlated multiexcitons in the {Bethe}-{Salpeter} equation},}\ }\href {\doibase 10.1103/6vsh-ck4q} {\bibfield  {journal} {\bibinfo  {journal} {Physical Review B}\ }\textbf {\bibinfo {volume} {112}},\ \bibinfo {pages} {115153} (\bibinfo {year} {2025})}\BibitemShut {NoStop}%
\bibitem [{\citenamefont {Fetter}\ and\ \citenamefont {Walecka}(2012)}]{fetter_quantum_2012}%
  \BibitemOpen
  \bibfield  {author} {\bibinfo {author} {\bibfnamefont {A.~L.}\ \bibnamefont {Fetter}}\ and\ \bibinfo {author} {\bibfnamefont {J.~D.}\ \bibnamefont {Walecka}},\ }\href@noop {} {\emph {\bibinfo {title} {Quantum {Theory} of {Many}-{Particle} {Systems}}}},\ Dover {Books} on {Physics}\ (\bibinfo  {publisher} {Dover Publications},\ \bibinfo {address} {Newburyport},\ \bibinfo {year} {2012})\BibitemShut {NoStop}%
\bibitem [{\citenamefont {Romaniello}, \citenamefont {Guyot},\ and\ \citenamefont {Reining}(2009)}]{Romaniello_2009}%
  \BibitemOpen
  \bibfield  {author} {\bibinfo {author} {\bibfnamefont {P.}~\bibnamefont {Romaniello}}, \bibinfo {author} {\bibfnamefont {S.}~\bibnamefont {Guyot}}, \ and\ \bibinfo {author} {\bibfnamefont {L.}~\bibnamefont {Reining}},\ }\bibfield  {title} {\enquote {\bibinfo {title} {The self-energy beyond {$GW$}: Local and nonlocal vertex corrections},}\ }\href {\doibase 10.1063/1.3249965} {\bibfield  {journal} {\bibinfo  {journal} {The Journal of Chemical Physics}\ }\textbf {\bibinfo {volume} {131}},\ \bibinfo {pages} {154111} (\bibinfo {year} {2009})}\BibitemShut {NoStop}%
\bibitem [{\citenamefont {Neuhauser}, \citenamefont {Rabani},\ and\ \citenamefont {Baer}(2012)}]{Neuhauser2012}%
  \BibitemOpen
  \bibfield  {author} {\bibinfo {author} {\bibfnamefont {D.}~\bibnamefont {Neuhauser}}, \bibinfo {author} {\bibfnamefont {E.}~\bibnamefont {Rabani}}, \ and\ \bibinfo {author} {\bibfnamefont {R.}~\bibnamefont {Baer}},\ }\bibfield  {title} {\enquote {\bibinfo {title} {Expeditious stochastic approach for {MP2} energies in large electronic systems},}\ }\href {\doibase 10.1021/ct300946j} {\bibfield  {journal} {\bibinfo  {journal} {Journal of Chemical Theory and Computation}\ }\textbf {\bibinfo {volume} {9}},\ \bibinfo {pages} {24--27} (\bibinfo {year} {2012})}\BibitemShut {NoStop}%
\bibitem [{\citenamefont {Baer}, \citenamefont {Neuhauser},\ and\ \citenamefont {Rabani}(2013)}]{PhysRevLett.111.106402}%
  \BibitemOpen
  \bibfield  {author} {\bibinfo {author} {\bibfnamefont {R.}~\bibnamefont {Baer}}, \bibinfo {author} {\bibfnamefont {D.}~\bibnamefont {Neuhauser}}, \ and\ \bibinfo {author} {\bibfnamefont {E.}~\bibnamefont {Rabani}},\ }\bibfield  {title} {\enquote {\bibinfo {title} {{Self-Averaging Stochastic Kohn-Sham Density-Functional Theory}},}\ }\href {\doibase 10.1103/PhysRevLett.111.106402} {\bibfield  {journal} {\bibinfo  {journal} {Phys. Rev. Lett.}\ }\textbf {\bibinfo {volume} {111}},\ \bibinfo {pages} {106402} (\bibinfo {year} {2013})}\BibitemShut {NoStop}%
\bibitem [{\citenamefont {Bradbury}\ \emph {et~al.}(2023{\natexlab{b}})\citenamefont {Bradbury}, \citenamefont {Allen}, \citenamefont {Nguyen},\ and\ \citenamefont {Neuhauser}}]{bradbury_deterministicfragmented-stochastic_2023}%
  \BibitemOpen
  \bibfield  {author} {\bibinfo {author} {\bibfnamefont {N.~C.}\ \bibnamefont {Bradbury}}, \bibinfo {author} {\bibfnamefont {T.}~\bibnamefont {Allen}}, \bibinfo {author} {\bibfnamefont {M.}~\bibnamefont {Nguyen}}, \ and\ \bibinfo {author} {\bibfnamefont {D.}~\bibnamefont {Neuhauser}},\ }\bibfield  {title} {\enquote {\bibinfo {title} {Deterministic/{Fragmented}-{Stochastic} {Exchange} for {Large}-{Scale} {Hybrid} {DFT} {Calculations}},}\ }\href {\doibase 10.1021/acs.jctc.3c00987} {\bibfield  {journal} {\bibinfo  {journal} {Journal of Chemical Theory and Computation}\ }\textbf {\bibinfo {volume} {19}},\ \bibinfo {pages} {9239--9247} (\bibinfo {year} {2023}{\natexlab{b}})}\BibitemShut {NoStop}%
\bibitem [{\citenamefont {Gao}\ \emph {et~al.}(2015)\citenamefont {Gao}, \citenamefont {Neuhauser}, \citenamefont {Baer},\ and\ \citenamefont {Rabani}}]{gao_sublinear_2015}%
  \BibitemOpen
  \bibfield  {author} {\bibinfo {author} {\bibfnamefont {Y.}~\bibnamefont {Gao}}, \bibinfo {author} {\bibfnamefont {D.}~\bibnamefont {Neuhauser}}, \bibinfo {author} {\bibfnamefont {R.}~\bibnamefont {Baer}}, \ and\ \bibinfo {author} {\bibfnamefont {E.}~\bibnamefont {Rabani}},\ }\bibfield  {title} {\enquote {\bibinfo {title} {Sublinear scaling for time-dependent stochastic density functional theory},}\ }\href {\doibase 10.1063/1.4905568} {\bibfield  {journal} {\bibinfo  {journal} {The Journal of Chemical Physics}\ }\textbf {\bibinfo {volume} {142}},\ \bibinfo {pages} {034106} (\bibinfo {year} {2015})}\BibitemShut {NoStop}%
\bibitem [{\citenamefont {Allen}, \citenamefont {Nguyen},\ and\ \citenamefont {Neuhauser}(2024)}]{Allen2024}%
  \BibitemOpen
  \bibfield  {author} {\bibinfo {author} {\bibfnamefont {T.}~\bibnamefont {Allen}}, \bibinfo {author} {\bibfnamefont {M.}~\bibnamefont {Nguyen}}, \ and\ \bibinfo {author} {\bibfnamefont {D.}~\bibnamefont {Neuhauser}},\ }\bibfield  {title} {\enquote {\bibinfo {title} {{$GW$} with hybrid functionals for large molecular systems},}\ }\href {\doibase 10.1063/5.0219839} {\bibfield  {journal} {\bibinfo  {journal} {The Journal of Chemical Physics}\ }\textbf {\bibinfo {volume} {161}},\ \bibinfo {pages} {114116} (\bibinfo {year} {2024})}\BibitemShut {NoStop}%
\bibitem [{\citenamefont {Thomas}\ \emph {et~al.}(2026)\citenamefont {Thomas}, \citenamefont {Nguyen}, \citenamefont {Bazile}, \citenamefont {Allen}, \citenamefont {Li}, \citenamefont {Li}, \citenamefont {Del~Ben}, \citenamefont {Deslippe},\ and\ \citenamefont {Neuhauser}}]{Thomas2026}%
  \BibitemOpen
  \bibfield  {author} {\bibinfo {author} {\bibfnamefont {P.~S.}\ \bibnamefont {Thomas}}, \bibinfo {author} {\bibfnamefont {M.}~\bibnamefont {Nguyen}}, \bibinfo {author} {\bibfnamefont {D.}~\bibnamefont {Bazile}}, \bibinfo {author} {\bibfnamefont {T.}~\bibnamefont {Allen}}, \bibinfo {author} {\bibfnamefont {B.~Y.}\ \bibnamefont {Li}}, \bibinfo {author} {\bibfnamefont {W.}~\bibnamefont {Li}}, \bibinfo {author} {\bibfnamefont {M.}~\bibnamefont {Del~Ben}}, \bibinfo {author} {\bibfnamefont {J.}~\bibnamefont {Deslippe}}, \ and\ \bibinfo {author} {\bibfnamefont {D.}~\bibnamefont {Neuhauser}},\ }\bibfield  {title} {\enquote {\bibinfo {title} {Stochastic {$GW$}-{GPU}: Rapid quasiparticle energies for molecules beyond 10,000 atoms},}\ }\href {\doibase 10.1021/acs.jctc.6c00116} {\bibfield  {journal} {\bibinfo  {journal} {Journal of Chemical Theory and Computation}\ }\textbf {\bibinfo {volume} {22}},\ \bibinfo {pages} {3960--3970} (\bibinfo {year} {2026})}\BibitemShut {NoStop}%
\end{thebibliography}%

\end{document}